\documentclass[journal]{IEEEtran}
 
\usepackage{cite}
\usepackage[nowarn,acronyms,nonumberlist,nopostdot,nomain,nogroupskip]{glossaries}
\usepackage{graphicx, subcaption}
\newacronym{3gpp}{3GPP}{3rd Generation Partnership Project}
\newacronym{5g}{5G}{5\textsuperscript{th} Generation}
\newacronym{5gc}{5GC}{5G Core}
\newacronym{bs}{BS}{Base Station}
\newacronym{abft}{A-BFT}{Association-BeamForming Training}
\newacronym[firstplural=Access Categories (ACs)]{ac}{AC}{Access Category}
\newacronym{adc}{ADC}{Analog to Digital Converter}
\newacronym{addts}{ADDTS}{Add Traffic Stream}
\newacronym{afbw}{AFBW}{Average Fading Bandwidth}
\newacronym{aid}{AID}{Association Identifier}
\newacronym{aimd}{AIMD}{Additive Increase Multiplicative Decrease}
\newacronym{am}{AM}{Acknowledged Mode}
\newacronym{amc}{AMC}{Adaptive Modulation and Coding}
\newacronym{ampdu}{A-MPDU}{MAC Protocol Data Unit Aggregation}
\newacronym{aoa}{AoA}{Angle of Arrival}
\newacronym{aod}{AoD}{Angle of Departure}
\newacronym{ap}{AP}{Access Point}
\newacronym{app}{APP}{Application Layer}
\newacronym{aqm}{AQM}{Active Queue Management}
\newacronym{ar}{AR}{Augmented Reality}
\newacronym{ati}{ATI}{Announcement Transmission Interval}
\newacronym{awgn}{AGWN}{Additive White Gaussian Noise}
\newacronym{awv}{AWV}{Antenna Weight Vector}
\newacronym{balia}{BALIA}{Balanced Link Adaptation}
\newacronym{bdp}{BDP}{Bandwidth-Delay Product}
\newacronym{bf}{BF}{Beamforming}
\newacronym{bhi}{BHI}{Beacon Header Interval}
\newacronym{bi}{BI}{Beacon Interval}
\newacronym{brp}{BRP}{Beam Refinement Protocol}
\newacronym{bss}{BSS}{Basic Service Set}
\newacronym{bti}{BTI}{Beacon Transmission Interval}
\newacronym{cad}{CAD}{Computer-aided Design}
\newacronym{cbap}{CBAP}{Contention-Based Access Period}
\newacronym{cbr}{CBR}{Constant Bitrate}
\newacronym{cc}{CC}{Congestion Control}
\newacronym{cdf}{CDF}{Cumulative Distribution Function}
\newacronym{cir}{CIR}{Channel Impulse Response}
\newacronym{cn}{CN}{Core Network}
\newacronym{cp}{CP}{Control Plane}
\newacronym{cqi}{CQI}{Channel Quality Indicator}
\newacronym{crs}{CRS}{Cell Reference Signal}
\newacronym{csirs}{CSI-RS}{Channel State Information - Reference Signal}
\newacronym{csmaca}{CSMA/CA}{Carrier Sense Multiple Access with Collision Avoidance}
\newacronym{cts}{CTS}{Clear to Send}
\newacronym{d2d}{D2D}{Device-to-device}
\newacronym{dc}{DC}{Dual Connectivity}
\newacronym{dce}{DCE}{Direct Code Execution}
\newacronym{dcf}{DCF}{Distributed Coordination Function}
\newacronym{dci}{DCI}{Downlink Control Information}
\newacronym{delts}{DELTS}{Delete Traffic Stream}
\newacronym{dked}{DKED}{Double Knife Edge Diffraction}
\newacronym{dl}{DL}{Downlink}
\newacronym{dmg}{DMG}{Directional Multi-Gigabit}
\newacronym{dmr}{DMR}{Deadline Miss Ratio}
\newacronym{dmrs}{DMRS}{DeModulation Reference Signal}
\newacronym{dti}{DTI}{Data Transmission Interval}
\newacronym{dtmke}{DTMKE}{Double-truncated Multiple Knife-edge}
\newacronym{e2e}{E2E}{End-to-End}
\newacronym{ecn}{ECN}{Explicit Congestion Notification}
\newacronym{edca}{EDCA}{Enhanced Distributed Channel Access}
\newacronym{edf}{EDF}{Earliest Deadline First}
\newacronym{enb}{eNB}{evolved Node Base}
\newacronym{endc}{EN-DC}{E-UTRAN-\gls{nr} \gls{dc}}
\newacronym{epc}{EPC}{Evolved Packet Core}
\newacronym{es}{ES}{Edge Server}
\newacronym{ese}{ESE}{Extended Schedule Element}
\newacronym{fdd}{FDD}{Frequency Division Duplexing}
\newacronym{fdma}{FDMA}{Frequency Division Multiple Access}
\newacronym{fov}{FoV}{Field-of-View}
\newacronym{fs}{FS}{Fast Switching}
\newacronym{ftp}{FTP}{File Transfer Protocol}
\newacronym{fwa}{FWA}{Fixed Wireless Access}
\newacronym{gnb}{gNB}{Next Generation Node Base}
\newacronym{harq}{HARQ}{Hybrid Automatic Repeat reQuest}
\newacronym{hetnet}{HetNet}{Heterogeneous Network}
\newacronym{hh}{HH}{Hard Handover}
\newacronym{hol}{HOL}{Head-of-Line}
\newacronym{hqf}{HQF}{Highest-quality-first}
\newacronym{ia}{IA}{Initial Access}
\newacronym{iab}{IAB}{Integrated Access and Backhaul}
\newacronym{ibss}{IBSS}{Independent Basic Service Set}
\newacronym{id}{ID}{Identifier}
\newacronym{imt}{IMT}{International Mobile Telecommunication}
\newacronym{inr}{INR}{Interference to Noise Ratio}
\newacronym{iot}{IoT}{Internet of Things}
\newacronym{ipa}{IPA}{Inter-Packet Arrival}
\newacronym{ism}{ISM}{Industrial, Scientific, and Medical}
\newacronym{kpi}{KPI}{Key Performance Indicator}
\newacronym{lcf}{LCF}{Level Crossing Frequency}
\newacronym{lcr}{LCR}{Level Crossing Rate}
\newacronym{los}{LoS}{Line-of-Sight}
\newacronym{lp}{LP}{Low Power}
\newacronym{lte}{LTE}{Long Term Evolution}
\newacronym{m2m}{M2M}{Machine to Machine}
\newacronym{mac}{MAC}{Medium Access Control}
\newacronym{mc}{MC}{Multi-Connectivity}
\newacronym{mcs}{MCS}{Modulation and Coding Scheme}
\newacronym{mec}{MEC}{Mobile Edge Cloud}
\newacronym{mi}{MI}{Mutual Information}
\newacronym{mib}{MIB}{Master Information Block}
\newacronym{mimo}{MIMO}{Multiple Input, Multiple Output}
\newacronym{mumimo}{MU-MIMO}{Multi-User Multiple Input, Multiple Output}
\newacronym{ml}{ML}{Machine Learning}
\newacronym{mlr}{MLR}{Maximum-local-rate}
\newacronym[plural=\gls{mme}s,firstplural=Mobility Management Entities (MMEs)]{mme}{MME}{Mobility Management Entity}
\newacronym{mmwave}{mmWave}{millimeter wave}
\newacronym{moi}{MoI}{Method of Images}
\newacronym{mpc}{MPC}{Multi Path Component}
\newacronym{mptcp}{MPTCP}{Multipath TCP}
\newacronym{mr}{MR}{Maximum Rate}
\newacronym{mrdc}{MR-DC}{Multi \gls{rat} \gls{dc}}
\newacronym{mss}{MSS}{Maximum Segment Size}
\newacronym{mtd}{MTD}{Machine-Type Device}
\newacronym{mtu}{MTU}{Maximum Transmission Unit}
\newacronym{nav}{NAV}{Network Allocation Vector}
\newacronym{ncbr}{NCBR}{Non-Constant Bitrate}
\newacronym{nfv}{NFV}{Network Function Virtualization}
\newacronym{nlos}{nLoS}{Non-Line-of-Sight}
\newacronym{nr}{NR}{New Radio}
\newacronym{nrmse}{NRMSE}{Normalized Root Mean Square Error}
\newacronym{ns3}{ns-3}{Network Simulator 3}
\newacronym{nsa}{NSA}{Non Standalone}
\newacronym{o2i}{O2I}{Outdoor-to-Indoor}
\newacronym{ofdm}{OFDM}{Orthogonal Frequency Division Multiplexing}
\newacronym{pa}{PA}{Position-aware}
\newacronym{pan}{PAN}{Personal Area Network}
\newacronym{pas}{PAS}{Power Angular Spectrum}
\newacronym{pbch}{PBCH}{Physical Broadcast Channel}
\newacronym{pbss}{PBSS}{Personal Basic Service Set}
\newacronym{pci}{PCI}{Physical Cell Identity}
\newacronym{pcp}{PCP}{\gls{pbss} Central Point}
\newacronym{pcpap}{PCP/AP}{\acrlong{pcp}/\acrlong{ap}}
\newacronym{pdcch}{PDCCH}{Physical Downlonk Control Channel}
\newacronym{pdcp}{PDCP}{Packet Data Convergence Protocol}
\newacronym{pdsch}{PDSCH}{Physical Downlink Shared Channel}
\newacronym{pdu}{PDU}{Packet Data Unit}
\newacronym{pf}{PF}{Proportional Fair}
\newacronym{pgw}{PGW}{Packet Gateway}
\newacronym{phy}{PHY}{Physical Layer}
\newacronym{poc}{PoC}{Proof of Concept}
\newacronym{ppp}{PPP}{Poisson Point Process}
\newacronym{prb}{PRB}{Physical Resource Block}
\newacronym{pss}{PSS}{Primary Synchronization Signal}
\newacronym{pucch}{PUCCH}{Physical Uplink Control Channel}
\newacronym{pusch}{PUSCH}{Physical Uplink Shared Channel}
\newacronym{qd}{QD}{Quasi Deterministic}
\newacronym{qoe}{QoE}{Quality of Experience}
\newacronym{qos}{QoS}{Quality of Service}
\newacronym{rach}{RACH}{Random Access Channel}
\newacronym{ran}{RAN}{Radio Access Network}
\newacronym[firstplural=Radio Access Technologies (RATs)]{rat}{RAT}{Radio Access Technology}
\newacronym{red}{RED}{Random Early Detection}
\newacronym{rf}{RF}{Radio Frequency}
\newacronym{rl}{RL}{Reinforcement Learning}
\newacronym{rlc}{RLC}{Radio Link Control}
\newacronym{rlf}{RLF}{Radio Link Failure}
\newacronym{rma}{RMa}{Rural Macro}
\newacronym{rms}{RMS}{Root Mean Square}
\newacronym{rr}{RR}{Round Robin}
\newacronym{rrc}{RRC}{Radio Resource Control}
\newacronym{rrm}{RRM}{Radio Resource Management}
\newacronym{rs}{RS}{Remote Server}
\newacronym{rsrp}{RSRP}{Reference Signal Received Power}
\newacronym{rsrq}{RSRQ}{Reference Signal Received Quality}
\newacronym{rss}{RSS}{Received Signal Strength}
\newacronym{rssi}{RSSI}{Received Signal Strength Indicator}
\newacronym{rt}{RT}{Ray Tracer}
\newacronym{rts}{RTS}{Request to Send}
\newacronym{rtt}{RTT}{Round Trip Time}
\newacronym{rw}{RW}{Receive Window}
\newacronym{rx}{RX}{Receiver}
\newacronym{sa}{SA}{Standalone}
\newacronym{sack}{SACK}{Selective Acknowledgment}
\newacronym{sap}{SAP}{Service Access Point}
\newacronym{sc}{SC}{Single Carrier}
\newacronym{sch}{SCH}{Secondary Cell Handover}
\newacronym{scm}{SCM}{Spatial Channel Model}
\newacronym{scoot}{SCOOT}{Split Cycle Offset Optimization Technique}
\newacronym{sdma}{SDMA}{Spatial Division Multiple Access}
\newacronym{sdr}{SDR}{Software Defined Radio}
\newacronym{si}{SI}{Study Item}
\newacronym{sib}{SIB}{Secondary Information Block}
\newacronym{sinr}{SINR}{Signal-to-Interference-plus-Noise Ratio}
\newacronym{sir}{SIR}{Signal-to-Interference Ratio}
\newacronym{sls}{SLS}{Sector-Level Sweep}
\newacronym{sm}{SM}{Saturation Mode}
\newacronym{snr}{SNR}{Signal-to-Noise Ratio}
\newacronym{son}{SON}{Self-Organizing Network}
\newacronym{sp}{SP}{Service Period}
\newacronym{spr}{SPR}{Service Period Request}
\newacronym{srs}{SRS}{Sounding Reference Signal}
\newacronym{ss}{SS}{Synchronization Signal}
\newacronym{ssb}{SSB}{Synchronization Signal Block}
\newacronym{ssrsrp}{SS-RSRP}{Synchronization Signal Reference Signal Received Power}
\newacronym{sss}{SSS}{Secondary Synchronization Signal}
\newacronym{ssw}{SSW}{Sector Sweep}
\newacronym{sta}{STA}{Station}
\newacronym{stb}{STB}{Set Top Box}
\newacronym{tb}{TB}{Transport Block}
\newacronym{tcp}{TCP}{Transmission Control Protocol}
\newacronym{tdd}{TDD}{Time Division Duplexing}
\newacronym{tdma}{TDMA}{Time Division Multiple Access}
\newacronym{tfl}{TfL}{Transport for London}
\newacronym{tgad}{TGad}{Task Group ad}
\newacronym{tgay}{TGay}{Task Group ay}
\newacronym{tsconst}{TSCONST}{Traffic Scheduling Constraint}
\newacronym{tm}{TM}{Transparent Mode}
\newacronym{trp}{TRP}{Transmitter Receiver Pair}
\newacronym{ts}{TS}{Traffic Stream}
\newacronym{tspec}{TSPEC}{Traffic Specification}
\newacronym{tti}{TTI}{Transmission Time Interval}
\newacronym{ttt}{TTT}{Time-to-Trigger}
\newacronym{tx}{TX}{Transmitter}
\newacronym[firstplural=Transmission Opportunities (TXOPs)]{txop}{TXOP}{Transmission Opportunity}
\newacronym{udp}{UDP}{User Datagram Protocol}
\newacronym{ue}{UE}{User Equipment}
\newacronym{ul}{UL}{Uplink}
\newacronym{ula}{ULA}{Uniform Linear Array}
\newacronym{um}{UM}{Unacknowledged Mode}
\newacronym{uma}{UMa}{Urban Macro}
\newacronym{umi}{UMi}{Urban Micro}
\newacronym{uml}{UML}{Unified Modeling Language}
\newacronym{utc}{UTC}{Urban Traffic Control}
\newacronym{v2v}{V2V}{Vehicle-to-Vehicle}
\newacronym{vbr}{VBR}{Variable Bit Rate}
\newacronym{vm}{VM}{Virtual Machine}
\newacronym{vr}{VR}{Virtual Reality}
\newacronym{wbf}{WBF}{Wired Bias Function}
\newacronym{wf}{WF}{Wired-first}
\newacronym{wifi}{Wi-Fi}{Wireless Fidelity}
\newacronym{wigig}{WiGig}{Wireless Gigabit}
\newacronym{wlan}{WLAN}{Wireless Local Area Network}
\newacronym{ber}{BER}{Bit Error Rate}
\newacronym{arf}{ARF}{Auto Rate Fallback}
\newacronym{semm}{SEMM}{SPCA-EDCA Mixed Mode}
\newacronym{ppdu}{PPDU}{PHY Protocol Data Unit}
\newacronym{sla}{SLA}{Service-Level Agreement}
\newacronym{re}{RE}{Resource Element}
\usepackage{multirow}
\usepackage[hyphens]{url}
\usepackage{amsmath,amssymb,amsfonts}
\usepackage{algorithmic}
\usepackage{graphicx}
\usepackage{textcomp}

\usepackage[font=small]{caption}
\usepackage{caption}

\usepackage[capitalize]{cleveref}
\crefname{figure}{Fig.}{Figs.}
\crefname{section}{Sec.}{Secs.}
\crefname{table}{Tab.}{Tabs.}

\usepackage{xcolor, soul}
\sethlcolor{green}

\def\BibTeX{{\rm B\kern-.05em{\sc i\kern-.025em b}\kern-.08em
    T\kern-.1667em\lower.7ex\hbox{E}\kern-.125emX}}
\begin{document}
\bstctlcite{IEEEexample:BSTcontrol}

\title{Sectors, Beams and Environmental Impact on the Performance of Commercial 5G mmWave Cells: an Empirical Study}

\author{Salman Mohebi$^{*}$, ~\IEEEmembership{Member,~IEEE,} Foivos Michelinakis$^{*}$, Ahmed Elmokashfi, Ole Grøndalen,\\ Kashif Mahmood, Andrea Zanella, ~\IEEEmembership{Senior member,~IEEE}

\thanks{
$^*$The first and second authors contributed equally to this paper.

This work has been supported by the European Community through the 5G-VINNI project (grant no. 815279) within the H2020-ICT-17-2017 research and innovation program, it also partially supported by the EU H2020 project ``WindMill,'' under the MSC Grant Agreement No. 813999.

Salman Mohebi and Andrea Zanella were with the Department of Information Engineering, University of Padova, Padova, Italy. Foivos Michelinakis and Ahmed Elmokashfi were with Simula Metropolitan, Oslo, Norway.  Ole Grøndalen and Kashif Mahmood were with Telenor Research, Oslo, Norway.
}
}

\maketitle

\begin{abstract}
\gls{mmwave} communication is one of the cornerstones of future generations of mobile networks.
While the performance of \gls{mmwave} links has been thoroughly investigated by simulations or testbeds, the behavior of this technology in real-world commercial setups has not yet been thoroughly documented.
In this paper, we address this gap and present the results of an empirical study to determine the actual performance of a commercial 5G \gls{mmwave} cell through on-field measurements.
We evaluate the signal and beam coverage map of an operational network as well as the end-to-end communication performance of a 5G \gls{mmwave} connection, considering various scenarios, including human body blockage effects, foliage-caused and rain-induced attenuation, and water surface effects. 
To the best of our knowledge, this paper is the first to report on a commercial deployment while not treating the radio as a black box.
Measurement results are compared with 3GPP's statistical channel models for \gls{mmwave} to check the possible gaps between simulated and actual performance.
This measurement analysis provides valuable information for researchers and 5G verticals to fully understand how a 5G \gls{mmwave} commercial access network operates in real-world, under various operational conditions, with buildings, humans, trees, water surfaces, etc.
\end{abstract}


\maketitle
\section{Introduction}\label{sec:intro}
\glsresetall

The abundant free spectrum available at \gls{mmwave} frequencies, spanning from 30~GHz to 300~GHz, makes \gls{mmwave} communication a key enabler for \gls{5g} systems to support bandwidth-hungry applications like online High Definition video streaming, augmented and virtual reality, and road-side vehicular communications.

However, transmission over \gls{mmwave} bands has its unique characteristics and adds new challenges, which are very different from those of sub-6~GHz communications.
In the last decade, a massive body of research has been carried out to
understand and model \glspl{mmwave}' propagation properties, mainly focusing on path-loss models, ray propagation mechanisms, material penetration, and atmospheric effects.

The first commercial 5G \gls{mmwave} systems have already been deployed in the last two years, and some early measurements~\cite{narayanan2020first,narayanan2021variegated,narayanan2020lumos5g} investigated the performance of these systems under various urban scenarios, revealing a high variability in the systems' performance, partially attributed to the high sensitivity to the propagation environment.
These studies are important because the commercial instalment may require to implement some changes, adaptations, and parameters’ setting that were not be required nor tested in \gls{poc} or pre-deployment  phases and that may potentially affect the system behavior in certain situations. 
Evidence of such a risk was reported, for example, in \cite{michelinakis2020dissecting} where the authors observed how an unexpected setting of some base station parameters had a dramatic impact of the energy consumption of narrowband-IoT nodes, significantly deviating from what predicted by models based on the protocol specifications. 
So, further research is required to fully understand the behavior of \glspl{mmwave} in an operational setup.
To this end, we have conducted a measurement campaign to analyze the impact of different environmental phenomena like rain, water surfaces, foliage, and human body blockage on the performance of an operational 5G \gls{mmwave} cell. 
We have also studied the signal coverage in different propagation environments for different sectors and beams. The purpose of this study is hence to understand to what extent the expected performance of \glspl{mmwave} is fulfilled in commercial settings, with all the complexity of an actual cellular system and of a real-world environment.
In many cases, our results confirm the system behavior already observed in previous studies based on non-commercial \gls{poc} deployments or predicted by theoretical models and simulations. 
However, in a few cases, we observed some nonconforming results that may be proxy of some problems in the deployment of the commercial solution. 

The analysis of the measurement results provides guidelines for planning future deployments and predicting the performance of 5G in different use cases, such as in case of fish farms/aquaculture~\cite{over_water}, or when the \glspl{ue} are located
inside forests or vegetated areas~\cite{vinni5g}, or when the \gls{los}
signal is blocked by buildings, moving objects or humans, as in dense urban environments~\cite{kaloxylos_alexandros_2020_3698113}.
Therefore, our observations are especially helpful to industries interested in deploying
5G over mmWave frequencies, but are not familiar with its intricacies.
 
In summary, our main contributions are as follows:
\begin{itemize}
    \item We present the coverage analysis of a commercial \gls{5g} mmWave cell by measuring the \gls{rsrp} in a complex real-world propagation environment (\Cref{sec:coverage_analysis}). 
    \item We then analyze the beam separation and gauge the difference with respect to the sector-level transmission (\Cref{sec:beam_separation}).
    \item We study different environmental impacts from the body and foliage blockage to rain and over-water transmission on \gls{mmwave} links on the commercial setup (\Cref{sec:measurement_result_analysis}).    
    \item We analyze the performance of \gls{nlos} \gls{mmwave} links in two different sectors, representing urban and suburban areas, observing that in an urban environment with multiple buildings and reflecting elements, the \gls{nlos} components of \gls{mmwave} signals can compensate for the lack of \gls{los} links, while this is not the case in the suburban environment (\Cref{sec:measurement_result_analysis}).
    \item We analyze the effect of the above-mentioned scenarios on the performance of end-to-end transmissions (\Cref{sec:measurement_result_analysis}).
    \item We compare the measurement results with \gls{3gpp}'s statistical channel models for the urban and rural environment both for omnidirectional \gls{rsrp} and transceiver's optimal antenna configuration, revealing the gap between the ideal simulated environment and the complex propagation environment (\Cref{sec:simulation}).
    \item Finally, we discuss how the above would affect real-world applications (\Cref{sec:result_summary}).
\end{itemize}

The remainder of this paper is structured as follows:
\Cref{sec:related_work} reviews the existing literature
around \gls{mmwave} propagation and early 5G \gls{mmwave} deployments. \Cref{sec:measurement_methodology} describes the  methodology used to conduct the measurement campaign. The observations and result analysis is provided in~\Cref{sec:measurement_result}, and \Cref{sec:simulation} compares the measurement results with some simple simulated scenarios. Finally, \Cref{sec:result_summary} concludes this article summarizing key findings and take-home messages.

\section{Related work}\label{sec:related_work}
\begin{table*}[]
\caption{Related works}\label{tbl:related_works}
\begin{tabular}{lllll}
\hline
Ref.                                                             & Year & Scenario                     & Methodology                               & Contribution                                                  \\ \hline
\cite{narayanan2020first}                                       & 2020 & 5G urban                     & Measurement (commercial)                  & Performance of end-to-end transmission in different scenarios \\ 
\cite{narayanan2021variegated}                                  & 2021 & 5G urban                     & Measurement (commercial)                  & Performance, power consumption, and application QoS           \\ 
\cite{narayanan2020lumos5g}                                     & 2020 & 5G urban                     & Measurement (commercial)                  & Factors affect 5G performance in UE-side                   \\ 
\cite{pimienta2020indoor}                       & 2020 & LoS indoor corridor          & Measurement (testbed) & Path loss model                                               \\ 
\cite{khalily2018indoor}                        & 2018 & LoS indoor office            & Measurement (testbed)                     & Path loss/ Large scale fading                                 \\ 
\cite{lin2019millimeter}                        & 2019 & LoS/nLoS indoor lecture hall & Measurement (testbed)                     & mmWaves channel characteristics                               \\ 
\cite{zhao2017channel}                          & 2017 & Urban                        & Simulation/Measurement                    & Channel modeling                                              \\ 
\cite{ko2017millimeter}                         & 2017 & Urban/Indoor                 & Measurement (testbed)                     & Spatio-temporal features of channel                           \\ 
\cite{hur2018feasibility}                       & 2018 & Urban                        & Measurement (testbed)                               & Feasibility of mobility                                       \\ 
\cite{zhang2020measurement}                     & 2020 & Foliage/Suburb               & Measurement (testbed)                               & Propagation characterization                                  \\ 
\cite{bas201728}                                & 2017 & Foliage/Suburb               & Measurement (testbed)                               & Propagation characterization                                  \\ 
\cite{ko201728}                                 & 2017 & Foliage/Suburb               & Measurement (testbed)                               & Foliage propagation model                                     \\ 
\cite{virk2019modeling}                         & 2019 & Body blockage                & Measurement (testbed)                               & Human blockage model                                          \\ 
\cite{alammouri2019hand}                        & 2019 & Body blockage                & Simulation                                & Hand grip impact                                              \\ 
\cite{huang2019rain}                            & 2019 & Rain                         & Measurement (testbed)                               & Rain attenuation                                              \\ 
\cite{al2020statistical}              & 2020 & Rain                         & Measurement (testbed)                              & Rain attenuation                                              \\ 
\cite{poorzare2020open, poorzare2020challenges} & 2020 & End-to-end urban             & Simulation                                & Performance of TCP in mmWave                                  \\ 
\cite{polese2017tcp, polese2017tcp2}            & 2017 & End-to-end urban             & Simulation                                & Performance of TCP in mmWave                                  \\ 
\cite{zhang2019will}                            & 2019 & High speed/Urban/Indoor      & Simulation                                & Performance of TCP in mmWave                                  \\ 
\hline
\end{tabular}
\end{table*}

In the past few years, several studies have experimentally investigated the behavior of \gls{mmwave} propagation in different scenarios and conditions: indoors~\cite{pimienta2020indoor, khalily2018indoor, lin2019millimeter}; urban environment~\cite{zhao2017channel, ko2017millimeter, hur2018feasibility}; suburban and vegetated area~\cite{zhang2020measurement, bas201728, ko201728}; human body blockage~\cite{virk2019modeling, alammouri2019hand} and rain-induced fading~\cite{al2020statistical, huang2019rain}.
Further,~\cite{ren2021survey, poorzare2020challenges, poorzare2020open,polese2017tcp, polese2017tcp2,zhang2019will} study end-to-end transmissions over \glspl{mmwave}.
\Cref{tbl:related_works} presents a summary of the related work.

The studies mainly aim to characterize the propagation of \gls{mmwave} signals in different environments and under various circumstances.
For example, the measurement study in~\cite{zhao2017channel} considered the outdoor 32~GHz microcells to extract and develop a \gls{mmwave} channel model. 
The empirical result is then compared and validated through simulation.
Ko~\emph{et~al.} ~\cite{ko2017millimeter} investigated the wideband directional channel characteristic of \glspl{mmwave} in both indoor and urban environments to model the spatio-temporal features of the communication channel.
In~\cite{hur2018feasibility}, the authors investigate, through a  measurement study, the feasibility of mobility for a typical vehicular speed in the urban environment.

The propagation of \glspl{mmwave} in suburban and vegetated environments, surrounded by lots of
foliage, is highly different from the urban and indoor scenarios.
This matter has already been considered in the literature, where a vast body of research studies the effects of foliage attenuation on \gls{mmwave} propagation.
A measurement study in~\cite{zhang2020measurement} analyzes and extracts large-scale and small-scale propagation properties of 5G \glspl{mmwave} in various vegetated environments with different types and density of vegetation.
A real-time channel sounder is used in~\cite{bas201728} to measure \gls{mmwave} \gls{los} and \gls{nlos} channel responses in a suburban area with lined trees.
The authors then use the measurement results to generate a foliage propagation model based on the ITU-R terrestrial model.

The propagation of \glspl{mmwave} can be highly affected by different phenomena like rain and human body blockage.
\gls{3gpp} has recognized human body as one of the main obstacles affecting \gls{mmwave} propagation and causing large radio channel variations.
Human body blockage has been considered and modeled in the literature, based on \gls{dked}, wedge, and cylinder models.
In \cite{virk2019modeling}, human body blockage is measured at 15~GHz, 28~GHz, and 60 GHz for 15 humans with different heights and weights. They model the body blockage as a \gls{dtmke} scheme and compare the calculated diffraction with existing models such as the absorbing double knife-edge model and the \gls{3gpp} human blockage model.

The attenuation caused by precipitation can not be neglected at \gls{mmwave} bands, where rain droplets can absorb \gls{mmwave} signals whose wavelengths (1~mm to 10~mm) is comparable to the size of a raindrop (a few millimeters).
Rain attenuation in the 21.8~GHz and 73.5~GHz bands, based on a one-year measurement campaign in tropical regions, is presented in \cite{al2020statistical}. 
In~\cite{huang2019rain}, Huang~\emph{et~al.} employ a custom-designed channel sounder for 25.84~GHz and 77.52~GHz frequencies to measure the rain-induced signal attenuation for short-range \gls{mmwave} links.
\begin{figure*}
\centering
  \centering
  \includegraphics[width=.92\linewidth]{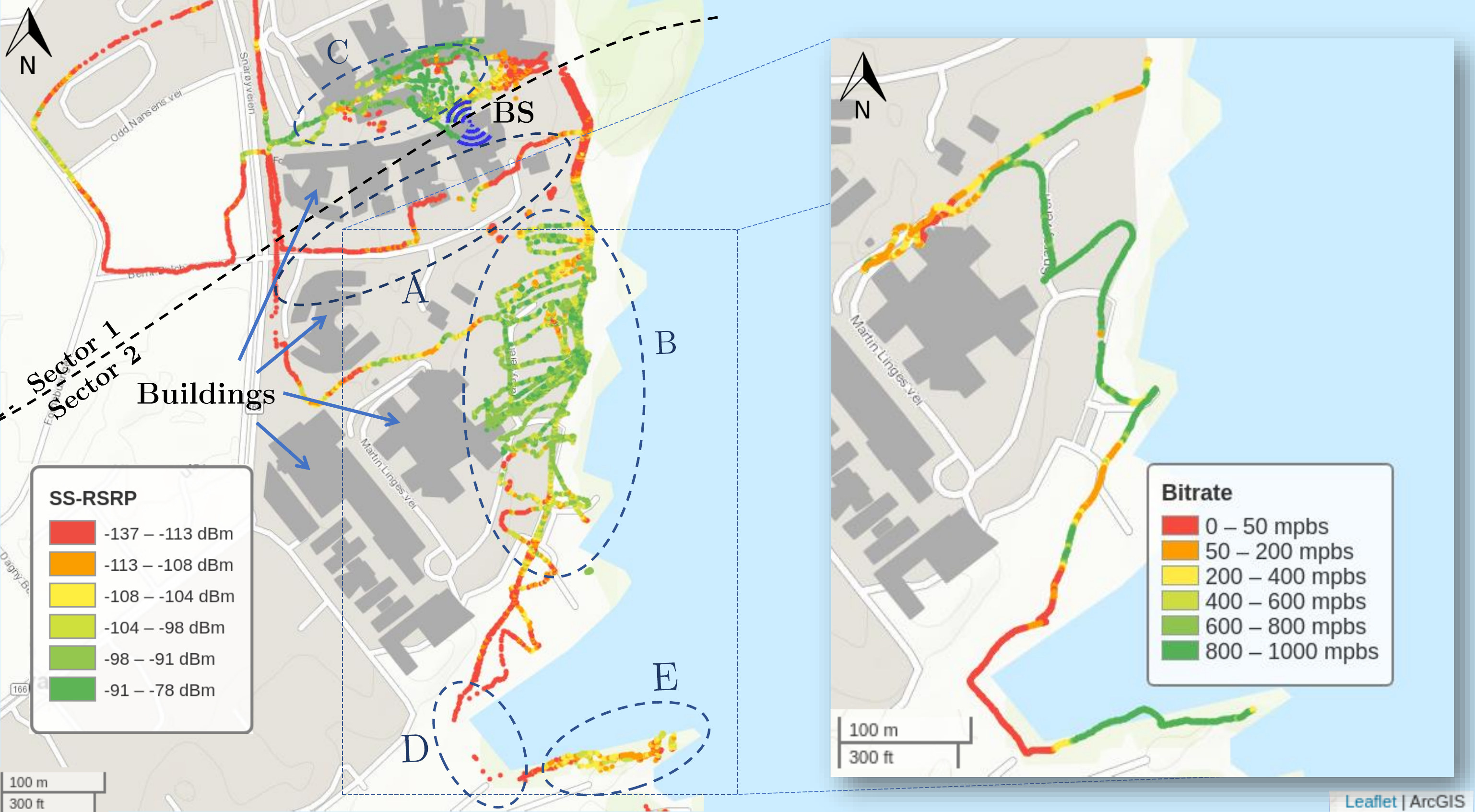}
\caption{Signal coverage map of the cell site, showing the maximum RSRP among all beams and PCIs. The zoomed-in area presents the bitrate.}
\label{fig:coverage_map}
\end{figure*}

From the user experience perspective, the efficiency of end-to-end transmission is critical when \gls{mmwave} links are part of the network, as the unsteady physical channel makes it difficult to support higher-layer connections.
In~\cite{poorzare2020challenges, poorzare2020open}, Poorzare \emph{et~al.}, presented an analysis of reliable end-to-end communications in 5G networks by investigating the effects of \gls{mmwave}  on \gls{tcp} performance and discussed the factors impacting the performance of 5G networks.
They further evaluated the performance of \gls{tcp} in urban environments under different conditions.
Polese~\emph{et~al.} evaluated the performance of \gls{tcp} over \gls{mmwave} links, relying on simulation~\cite{polese2017tcp, polese2017tcp2}.
They studied the behavior of multipath-\gls{tcp} on 28~GHz \gls{mmwave} links with a secondary \gls{lte} or 73~GHz \gls{mmwave} link.
Zhang~\emph{et~al.}~\cite{zhang2019will}, analyzed the performance of \gls{tcp} in \gls{mmwave} networks for high-speed \glspl{ue} in dense urban environments, where the \glspl{ue} are located at different geographical positions with \gls{los} and \gls{nlos} links to the \gls{bs} as well as indoor \glspl{ue}. They studied the performance of edge and remote servers as well as different \gls{tcp} variations.

Most of the previous works were conducted in test setups that were not equipped with commercial 5G \gls{mmwave} \glspl{bs} since commercial \gls{mmwave} deployments did not arrive until late 2019. 
Notable exceptions are studies by Narayanan~\emph{et~al.}~\cite{narayanan2020first,narayanan2021variegated,narayanan2020lumos5g}.
The study~\cite{narayanan2020first} presents a first look at the performance of two \gls{mmwave} and one mid-band commercial 5G deployments in US.
Using end-to-end performance measurements, Narayanan \emph{et~al.} tracked the interplay between propagation in the urban environment, blockage, and precipitation on applications performance.
They further expand their measurement campaign in~\cite{narayanan2021variegated} to include the power consumption and application \gls{qoe} of operational \gls{5g} networks by considering different deployment schemes, radio frequencies, protocol configurations, mobility patterns and upper-layer applications.
They also investigate the possibility of predicting network throughput in commercial \gls{mmwave} 5G networks~\cite{narayanan2020lumos5g}.
That work identified the different factors that affect 5G performance and proposed a context-aware throughput prediction framework based on Machine Learning techniques.

Like the work of Narayanan~\emph{et~al.}, we present an evaluation of a commercial 5G cell deployment.
However, to the best of our knowledge, this paper is the first to provide a fine-grained analysis of \gls{mmwave} propagation of a commercially deployed \gls{bs}. 
This offers concrete explanations for the main causes of performance degradation since we are not treating the radio as a black box.
We have also investigated a range of scenarios that are known to impact \gls{mmwave} propagation, including human body blockage, foliage, transmission over-water and rain.
While some of these have been investigated before, this paper is the first to analyze all of them in a commercial 5G \gls{mmwave} deployment with known parameters configuration.
We also note that this paper is the first to look at the effect of transmission over-water.
We further compare the measured \gls{rsrp} against \gls{3gpp}'s statistical \gls{mmwave} channel models for the urban and rural environments, considering both omnidirectional strongest \gls{rsrp} (transceiver's optimal antenna configuration), to study the simulation and actual performance (\Cref{sec:simulation}).

\section{Measurements Methodology} \label{sec:measurement_methodology}

This paper includes two measurement studies: the first study aims to analyze the coverage aspects of commercial \gls{5g} \gls{mmwave} cells, while the second study targets the end-to-end communication performance of a \gls{5g} network when \gls{mmwave} links are employed as part of the system.

The \gls{5g} \gls{mmwave} \gls{bs} is located on the roof of a 15~meters high building in Telenor's campus in Oslo, Norway.
The \gls{bs} is equipped with two Huawei HAAU5213 radio frequency units with 768 antenna arrays providing coverage to a northern and a southern sector as shown in \Cref{fig:coverage_map}.
Its frequency range is 26.5~GHz to 29.5~GHz with a maximum transmit power equal to 32.5~dBm.
It supports up to four carriers and can generate 16 different static beams, employing hybrid beamforming.
The black dashed line in the \Cref{fig:coverage_map} showcases the topological separation of the two sectors.
The northern sector (sector 1) points towards an open square surrounded by glass and steel buildings.
The southern sector (sector 2) is directed towards a peninsula with some buildings on the west and sea on the east.
Each sector has four 200~MHz wide channels (800~MHz total), with center frequencies between 26.6~GHz and 27.2~GHz.
We identify each channel by its respective \gls{pci},
where \glspl{pci} 101-104 belong to the northern sector and \glspl{pci} 301-304 to the southern. 
The operator can adjust the beams' boresight both in the horizontal and vertical plane. 
We did not have any control of the beams and no prior knowledge about their directions.
However, we were able to estimate the beams' directions based on the measurements we collected to create the coverage map from \gls{los} scenarios, if a single beam has the highest \gls{rsrp} in all locations of a measured area is considered to have an orientation towards this area.

\begin{figure}
  \centering
  \includegraphics[width=\linewidth]{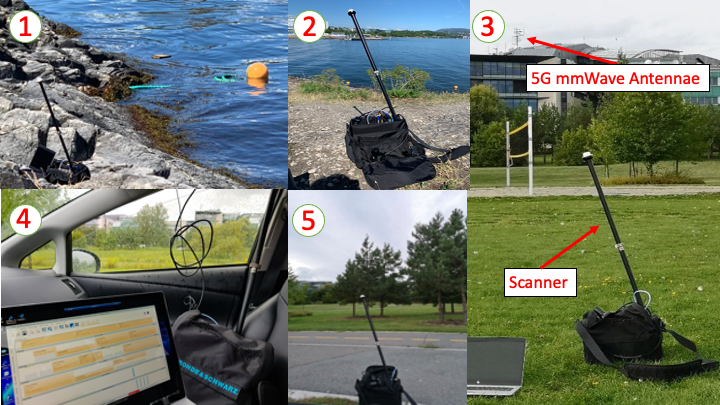}
  \caption{Some measurement locations: 1) close to water, 2) 6~m above water,
  3) Line of Sight, 4) rain, 5) foliage blockage.}
  \label{fig:measurement_setup}
\end{figure}

We collect channel quality information with a Rohde \& Schwarz scanner\cite{scanner} that can monitor all the relevant channels simultaneously using an omnidirectional antenna.
Note that our measurements do not consider the antenna gain that is expected in a commercial receiver (e.g., as smartphone). On the other hand, the isotropic antenna makes the measurements almost independent of the receiver orientation.
Since we are interested in how different factors affect propagation, the absence
of the receiver gain does not influence our evaluation.
The scanner was used to gather measurements across the site and under different conditions, collecting in total 535137 samples between April 2020 and September 2020.
Each sample contains several channel quality indicators, such as \gls{rsrp}, for all the detected \glspl{pci} and beams.
We create a coverage map by walking around the site with the scanner, and perform stationary measurements at selected locations, each lasting typically at least 5 minutes, to capture the time variations of the signal strength.
\Cref{fig:measurement_setup} shows the scanner and some of the measurement scenarios.

\begin{figure}
  \centering
  \includegraphics[width=0.95\linewidth]{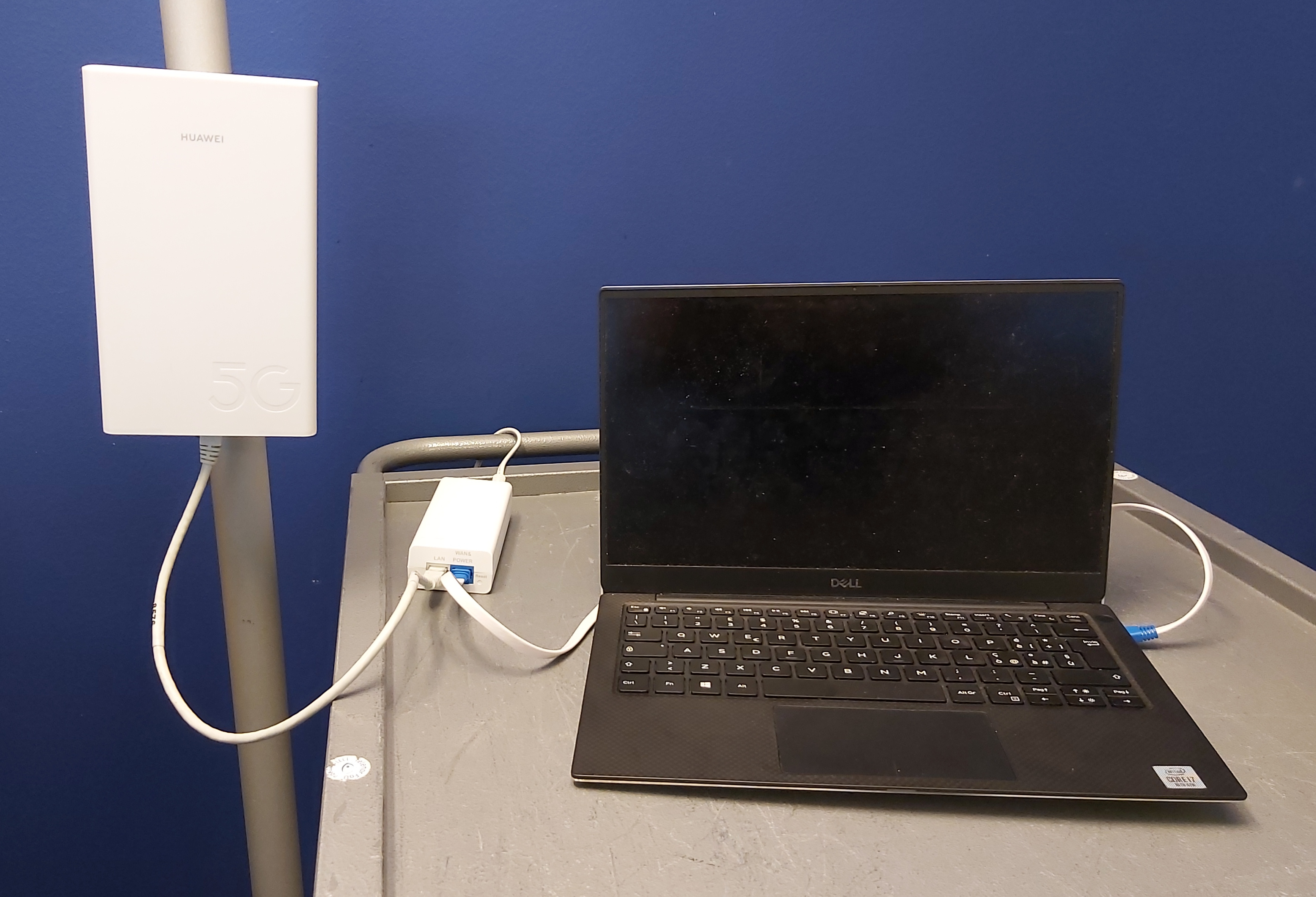}
  \caption{The measurement setup for the end-to-end communication experiment. The CPE is connected through a Gigabit Ethernet cable to a laptop running the scripts.}
  \label{fig:measurement_setup2}
\end{figure}

We also collected the measurements to analyze the performance of \gls{5g} \gls{mmwave} end-to-end transmission.
Because of \gls{bs} maintenance, the northern sector was not operational, and the measurements for the bitrate and delay study were done only in the southern sector. Each experiment was repeated
at exactly the same locations and with the same \gls{bs} configuration used for the channel quality measurements.
This experiment focused on the user experience, so we evaluated the end-to-end bitrate and \gls{rtt}.
The measurements were performed with a pre-production Huawei 5G CPE, supporting $2 \times 2$ MIMO and operating in \gls{nsa} mode, which can reach data rates of up to 1~Gbps.
A Gigabit Ethernet cable connects the CPE to a laptop which acts as the client.
Even though the \gls{bs} can achieve approximately 3~Gbps in downlink,
the Gigabit Ethernet limits the maximum transfer rate with the laptop to 1~Gbps. This does not pose an
obstacle for our analysis, since we are more interested on the cases where the network performance drops well below this limit, as a consequence of obstacles or other environmental phenomena.
\Cref{fig:measurement_setup2} shows the devices and the measurement setup used for this study.

The traffic sources are servers located inside the operator's network to avoid the effect of cross-traffic and congestion over the public Internet.
The delay performance was assessed from the \gls{rtt} measurements given by \texttt{ping}, with packets of 64 bytes (default setting) and of 1500 bytes (maximum size allowed by the Ethernet connection without requiring IP fragmentation).
Note that, to be transmitted over the wireless link, the bigger \texttt{ping} packets have to be split into multiple Transport Blocks when the system experiences bad signal, which results in the use of robust (but not very spectrum efficient) modulation and coding schemes.
The bitrate performance is evaluated through \texttt{iPerf3}\footnote{https://iperf.fr/}, a cross-platform tool for network performance measurement.
We use ten parallel \gls{tcp} connections, lasting at least 10 seconds, to get an estimation of the bitrate achieved at each measurement location.

\section{Measurement Results}\label{sec:measurement_result}
\subsection{Coverage analysis}\label{sec:coverage_analysis}
We first focus our analysis on the measured \gls{ssrsrp}, which
is the average power of the \glspl{re}
that carry the \gls{sss} transmitted within a \gls{ssb}~\cite{3gpp38215}.
The beams are time-multiplexed, thus there is no interference between the beams when the
\gls{ssrsrp} is measured.
For simplicity in the sequel we will refer to \gls{ssrsrp} as \gls{rsrp}.
Each \gls{ssb}/beam is assigned a unique number, called \gls{ssb} index.
Note that the values of the \gls{ssb} index were not contiguous.
Thus, in the subsequent plots, the numbering of \gls{ssb} indexes has gaps.

\Cref{fig:coverage_map} presents the \gls{mmwave} coverage map.
At each location, we were able to detect all the \glspl{pci} of the relevant sector and most of the beams.
Since a \gls{ue} would be attached to the dominant beam, i.e., that with the highest \gls{rsrp} among those detected by the \gls{ue}, in \Cref{fig:coverage_map} we report only this maximum \gls{rsrp} values.

In \gls{mmwave} bands, the \gls{rsrp} is dominated by the signal's \gls{los} components: missing these components can lead to significant attenuation. 
This can be easily seen in area \textbf{A} of \Cref{fig:coverage_map}, where the \gls{los} link is blocked by buildings and the \gls{rsrp} drops below -113~dBm.
Although the availability of \gls{los} components of a \gls{mmwave} signal is an important factor in determining coverage, other effects like signal diffraction, reflections from surrounded objects, and multipath propagation can compensate for their absence.
These effects are likely responsible for the relatively large \gls{rsrp} measured in Area \textbf{C} of sector 1, between the two buildings, where the \gls{los} link is blocked by the roof edge of the building hosting the \gls{bs}.
In contrast, in absence of reflecting or diffracting elements, there are no \gls{nlos} components of \gls{mmwave} signals.
This is the case of area \textbf{A}, where the signal propagates in a vegetated area without many reflecting elements, and of area \textbf{D}, where we did not record any significant \gls{rsrp} value at most of the locations. 

\begin{figure}[]
  \centering
  \includegraphics[width=0.9\linewidth]{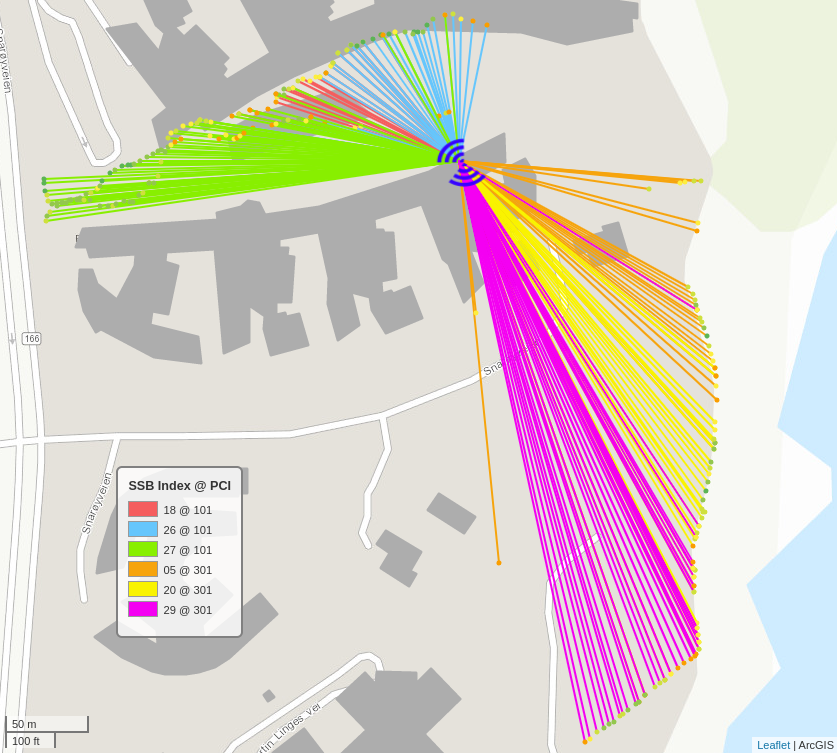}
  \caption{Dominant beams (lines) and the related RSRP values (points) in a subset of the locations.}
  \label{fig:beamCoverage}
\end{figure}

As expected, we did not observe a significant difference between \glspl{pci} for the same beam.
Moreover, in presence of \gls{los}, we did not record a strong dependency between signal power and distance to the \gls{bs}.
The attenuation due to the increasing distance is indeed marginal compared to the rest of the factors that affect the \gls{rsrp}, which fluctuates within a certain range as long as the receiver remains in the main lobe of the dominant beam. 
This behavior can be observed by considering a straight \gls{los} line in Area \textbf{B} and sampling locations at a distance from the \gls{bs} ranging from 200~m to 450~m. 
At such \gls{los} locations, the median \gls{rsrp} value of the dominant beam is always between -94~dBm and -100~dBm, regardless of the distance.
We hypothesize that it is the vertical antenna gain pattern that is causing this behavior.
At short distances we were located significantly below the boresight of the BS antenna, hence the antenna gain was low. As we moved further away we got closer to the boresight and the antenna gain increased. So the effects of larger pathloss and increasing antenna gain as the distance increased approximately  cancelled each other.
We discuss in more detail the relationship between \gls{rsrp} and distance in \cref{sec:simulation}, where \cref{fig:sim2} vizualises our measurements and compares
the maximum measured RSRP to state of the art models.

To construct the bitrate map, we launched 10 parallel continuous \gls{tcp} connections and walked around the cell site with normal speed, while tracking the location by external GPS.
Simultaneously, \texttt{tcpdump} captured the generated traffic.
We were able to get an estimate of the observed bitrate at each location by correlating the timestamps of the packet capture and the GPS log.
The packet capture is split into 100~ms bins and all the packets received during a bin are grouped together.
The bandwidth values are estimated by dividing the total number of bytes of all the packets in a bin by 100~ms.
Then, we assign this bandwidth value to the closest, by time, GPS entry.
As shown in \Cref{fig:coverage_map}, the bitrate in different locations is highly correlated with the \gls{rsrp}: the higher the \gls{rsrp}, the higher the bitrate.
As seen in the figure, in the \gls{los} locations, the maximum possible bitrate is achieved.
Even in area \textbf{E}, which is relatively far from the \gls{bs}, the bitrate is high.
Note that, the Gigabit Ethernet connecting the CPE to the Laptop is the bottleneck and limits the network speed to 1~Gbps.
In a very bad channel state (area \textbf{D}), \gls{tcp} still keeps the connection open but with a very low bitrate.

\subsection{Beam separation}
\label{sec:beam_separation}

\begin{figure}[]
  \centering
  \includegraphics[width=\linewidth]{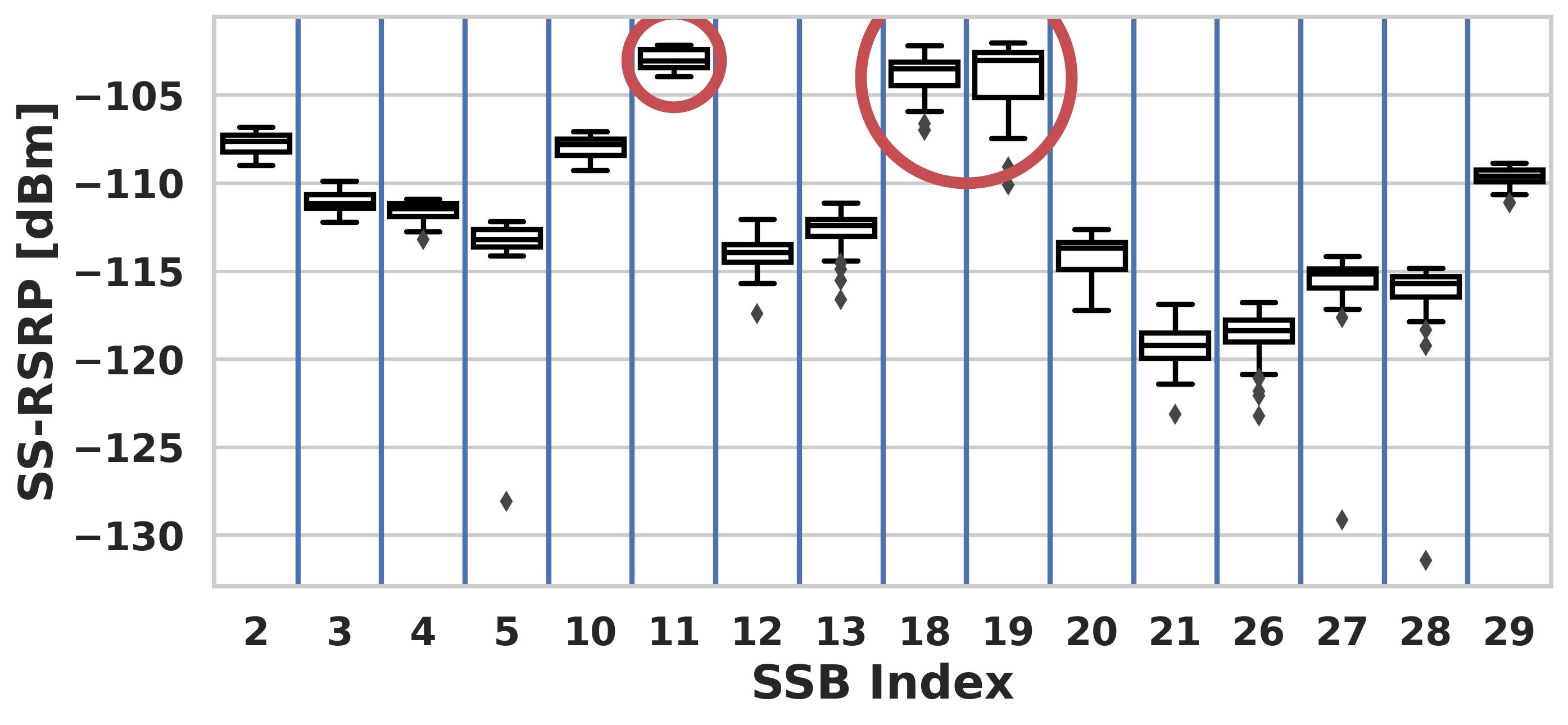}
  \caption{On rare occasions, we observe multiple dominant beams.
  Static measurements over a 5-minute period at a location 250~m away from
  the BS,  where several beams could be considered dominant (highlighted with red circles).}
  \label{fig:mulitpleDominantSSBs}
\end{figure}

A sharper beam can improve spatial separation between users, hence increasing
MU-MIMO performance, as well as reduce interference in multi-cell deployments.
To showcase beam separation, we select a few locations, creating a perimeter at the ground level around the \gls{bs}.
\Cref{fig:beamCoverage} color codes the dominant beam
at each location.
The beam lines drawn on the map are hypothetical, connecting the \gls{bs} and receiver location
and the actual beams are not as narrow as we have shown in the figure: signals from different beams can be detected at a much wider angle (side lobes) and even at the backside of the transmitter (back lobes).
According to~\cite{3gpp.38.304}, the \gls{rsrp} should be above -110~dBm to be detectable by 5G NR UEs, thus we filter out values below this threshold.
For each sector, we have also displayed only three out of 16 beams and have
not considered the beams that overlap in the vertical plane.
We observe a similar beam separation pattern across the vertical axis, by performing
measurements on several floors at the building opposite of the \gls{bs}
at sector 1.
The dominant beams at the ground level, third floor, fourth floor and roof are
different.

As commented in~\Cref{sec:coverage_analysis}, within a certain range, the actual distance between transmitter and receiver is not of much relevance as long as the receiver is within the main lobe of the dominant beam. Therefore, it is possible to get good signal even at long distances.
The furthest point from the \gls{bs} we could detect \gls{rsrp} higher than -110~dBm was 902~meters.
We can assume a commercial UE would be able to achieve an even bigger range, because of the receiver antenna gain.
At almost all the studied locations, a single beam had consistently and markedly higher \gls{rsrp} than the rest for the whole measurement duration, so beam selection was trivial.
However, it is possible to have multiple dominant beams in some locations, as shown in~\Cref{fig:mulitpleDominantSSBs}, where the \gls{rsrp} values for different beams are presented at a single location for one
\gls{pci} over a 5 minute period. We can observe that three beams, marked with red circles, have about the same median value.
The number of simultaneous beams is limited by the number of Transmit/Receive (transceiver) units in the \gls{bs}, so only one beam is transmitted in any given time/frequency slot for each \gls{pci}.
This time multiplexing avoids inter-beam interference, but beam selection becomes more complicated.
In such cases, it might be better to have a secondary criterion for beam selection, such as choosing the beam with the lowest standard deviation of \gls{rsrp}. 
Even more sophisticated beam selection algorithms~\cite{klautau2019lidar, zhu2020adaptive} might be required in a more complicated and dynamic propagation environment.
On the other hand, the slightly overlapped coverage regions of the \gls{ssb} beams are good for robustness (body blockage, moving cars, etc), where there is a higher chance of having at least a good beam at any time.
The other benefit of this slight overlap (or closely spaced beams) is to have a smooth user experience as a \gls{ue} moves from one beam’s coverage region to another beam's.

\subsection{Environmental impact on mmWave propagation}
\label{sec:measurement_result_analysis}

In the following, we analyze the effect of different environmental factors such as human body blockage, communication over-water surfaces, foliage and rain-induced attenuation on \gls{mmwave} propagation.      
\paragraph*{Human body blockage effects}

\begin{figure}[]
  \centering
  \includegraphics[width=\linewidth]{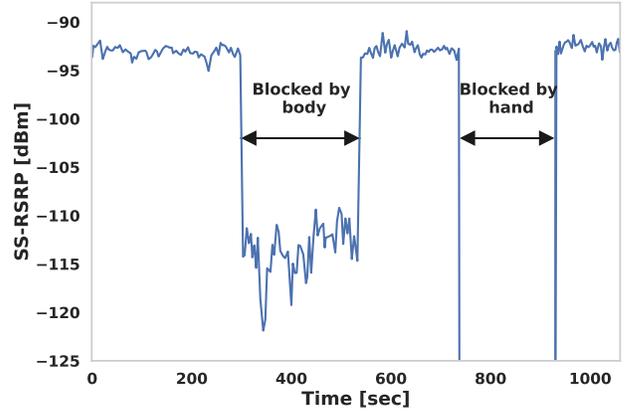}
  \caption{Body blockage effect on mmWave propagation at a location 260~m away from the BS with a LoS link.}
  \label{fig:body_blockage}
\end{figure}

\begin{table}[t]
\centering
\caption{Summary of the body blockage effect on bitrate.}
\begin{tabular}{lc}
    \hline
   Position & Bitrate [Mbps]  \\
    \hline
    Line of sight & 822  \\
    No line of sight (2 sitting 2~meters away from the CPE) &  613  \\
    No line of sight (2 standing right in front of the CPE) & 755  \\
    \hline
\end{tabular}
\label{tbl:bitrate_body_blockage}
\end{table}

To analyze the impact of human body blockage, we placed the scanner at a \gls{los} location 260~meters away from the \gls{bs}.
We then blocked the direct link from the \gls{bs} in two different ways.
First, by standing 10~cm away from the scanner, and later by folding the hand around the scanner's antenna, which are typical behaviors of smartphone users.
As shown in \Cref{fig:body_blockage}, we observed a 20-30~dB drop in \gls{rsrp} in the first experiment, which is in line with the literature~\cite{alammouri2019hand}.
In the latter case, all the signal components were removed, and we were unable to detect any signal.
In the first case, the received signal was likely due to the \gls{nlos} components reflected or scattered from the surrounding objects, or diffracted from the person were standing in front of the scanner.
Folding the hands around the scanner's antenna, instead, completely shielded the receiver from all the signal components, which explains the absence of significant \gls{rsrp} measurements.
The human body blockage effects on \gls{mmwave} communications have already been investigated and modeled in various ways, and the interested readers can refer to \cite{virk2019modeling, 3gpp.38.901} for more information.
Our experiments hence confirm this critical aspect also in commercial cell deployments. 

The effect of body blockage on bitrate is presented in~\Cref{tbl:bitrate_body_blockage}. 
We measured the bitrate with 10-parallel \gls{tcp} connections as above, while 1) two average-size people were sitting about 2~meters away from the CPE, completely covering the \gls{los} link, and 2) two people standing right in front of the CPE.
As seen in the table, the bitrate dropped from 822~Mbps to about 613~Mbps and 755~Mbps for the first and second scenarios, respectively.
This shows that despite the partial occlusion of \gls{los}, the performance at \gls{tcp} layer seems to remain acceptable.

\paragraph*{Foliage Attenuation}

\begin{figure}
\centering
\begin{subfigure}{0.8\linewidth}
  \centering
  \includegraphics[width=\linewidth]{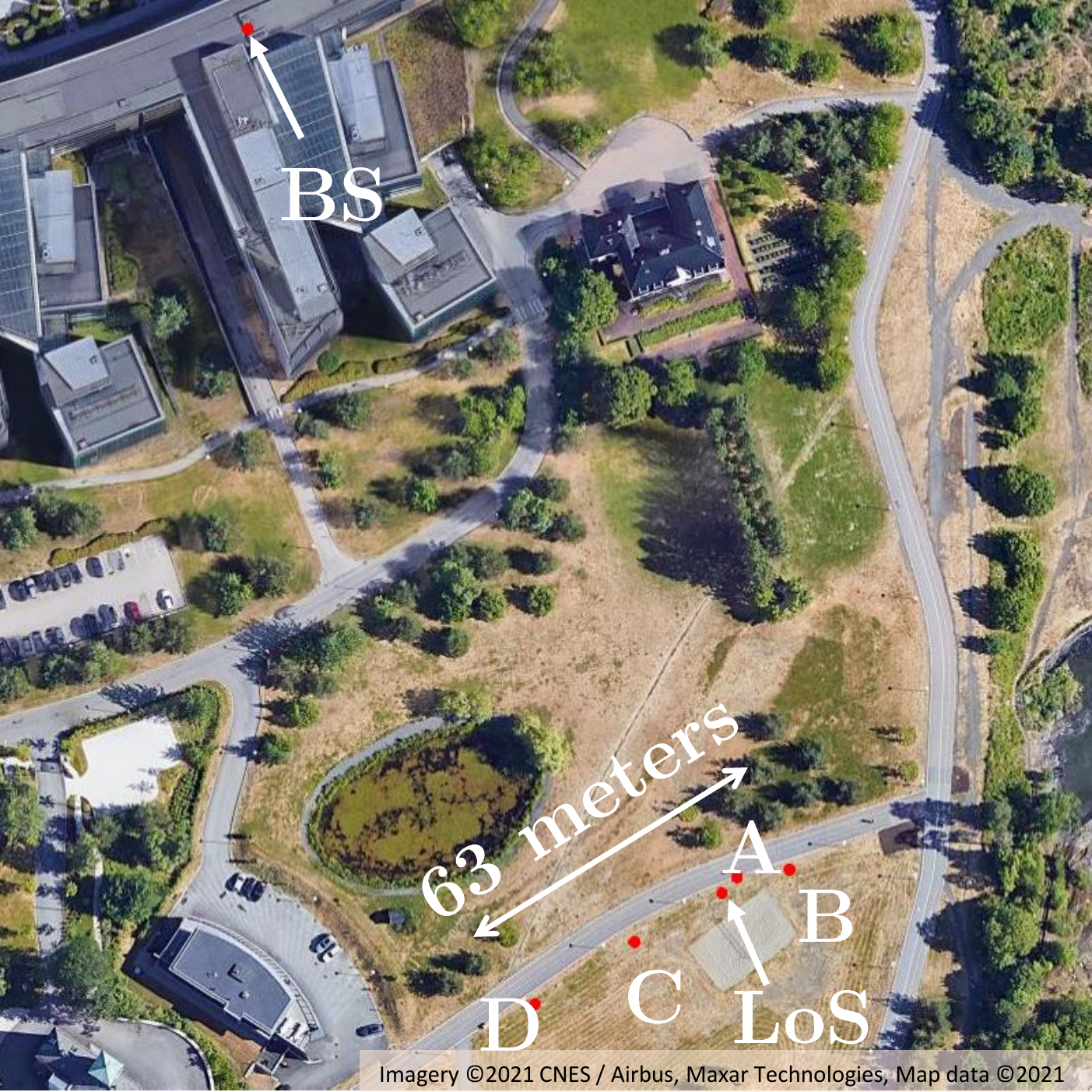}
\end{subfigure}
\begin{subfigure}{.9\linewidth}
  \centering
  \includegraphics[width=\linewidth]{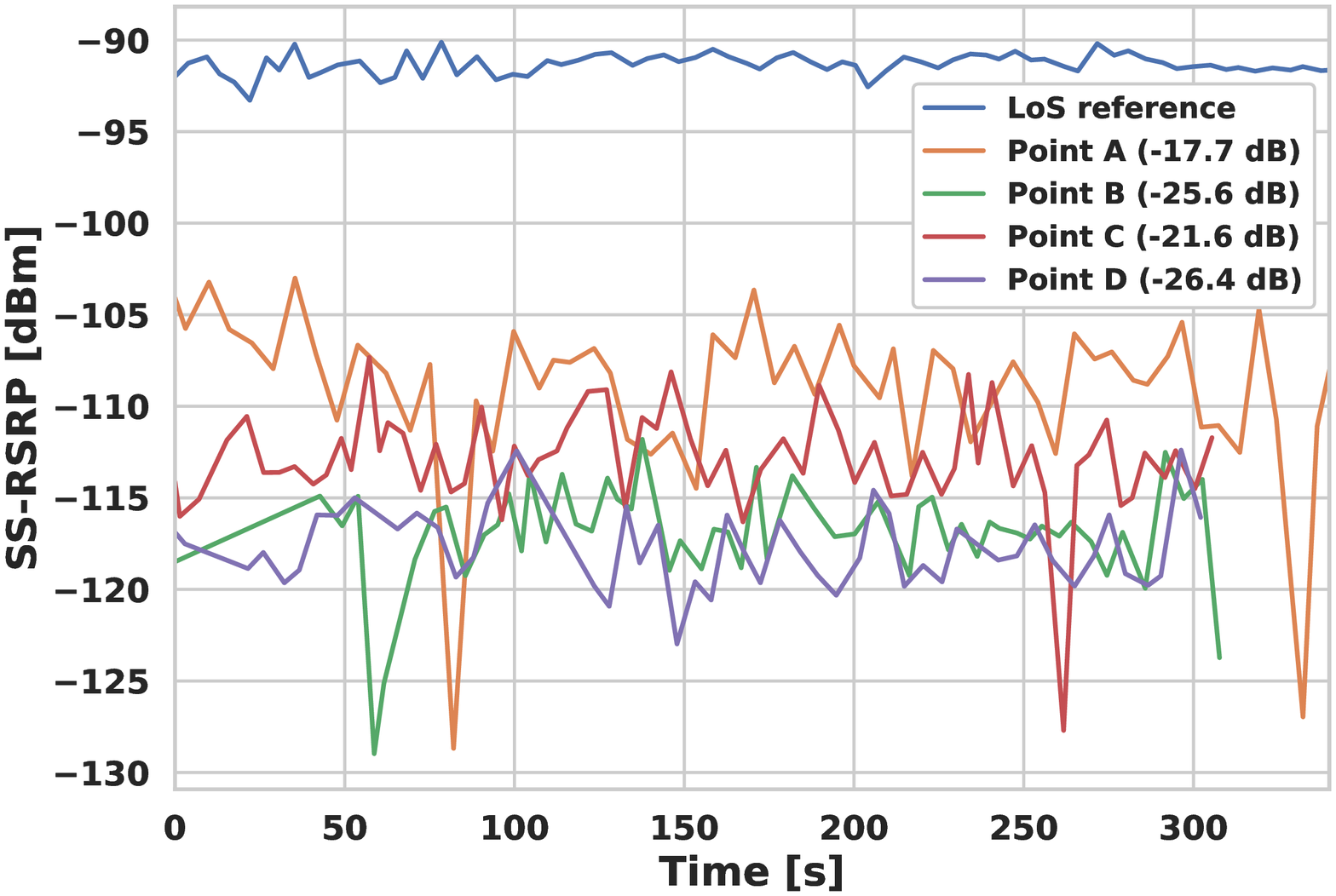}
\end{subfigure}\par\medskip
\caption{Foliage attenuation scenarios compared to \gls{los}. Locations A, B, C and D suffer from different types of foliage blockage. Distance from \gls{bs}: $\approx$220~m for all locations.}
\label{fig:treeWind}
\vspace{-.2cm}
\end{figure}

\begin{table}[t]
    \centering
    \caption{Summary of the effect of different types of foliage to bitrate and delay.}
    \begin{tabular}{llll}
    \hline
    Position & \begin{tabular}[c]{@{}c@{}}Bitrate\\ {[}Mbps{]} \end{tabular} & \begin{tabular}[c]{@{}c@{}}RTT\\ {[}ms{]} \end{tabular} & \begin{tabular}[c]{@{}c@{}}RTT big\\ {[}ms{]} \end{tabular} \\
    \hline
    LoS baseline besides the trees & 891 & 14.6 & 15.7 \\
    Besides a small tree & 831 & 13.4 & 16.6 \\
    Behind a single branch & 785 & 14.8 & 17.5 \\
    Behind big tree with sparse leaves & 451 & 14.8 & 19.6 \\
    Behind big tree & 323 & 15 & 17.8 \\
    
    \hline
    \end{tabular}
    \label{tbl:bitrate_delay_foliage}
\end{table}

The propagation characteristics of \gls{mmwave} frequencies in suburban and vegetated environments are very different from those in urban and indoor scenarios.
Foliage attenuation could significantly affect communication over this frequency range and should be considered in network planning for such suburban areas~\cite{zhang2020measurement}.
We carried out a set of measurements to investigate the effect of blocking the path between transmitter and receiver by vegetation and trees.
During the measurements, nearby weather stations reported wind intensity between 4.7 and 5.9~m/s, which is considered a ``gentle breeze" according to the Beaufort scale. The leaves and the twigs on small trees were in constant motion.

\Cref{fig:treeWind} shows the \gls{rsrp} at 4 different locations where the \gls{los} path is blocked by different types of trees.
The recorded \gls{rsrp} exhibits significant attenuation at all four locations, typically 15~dB to 30~dB lower than our measurements at a nearby reference \gls{los} location.
The intensity of attenuation highly depends on the type and shape of the blocking trees, with bigger trees causing higher attenuation.
As \Cref{fig:treeWind} illustrates, the smaller trees at location \textbf{A} attenuate the signal by 17.7~dB, while the bigger ones at location \textbf{D} decrease the received signal power by up to 26.4~dB.
This result is comparable with similar studies, where the authors reported $22.48\pm 0.92$~dB foliage attenuation in 26.5~GHz~\cite{rahim2017foliage}.
The high variations in \gls{rsrp} are likely due to wind that constantly moved trees' branches and leaves, which resulted in varying blockage and reflection patterns.

\Cref{tbl:bitrate_delay_foliage} summarizes the result for the foliage effect on bitrate and delay.
The measurements take place under a gentle breeze (wind speed between 2.4 to 5.1~m/s).
As was expected, the bitrate drop and delay are highly correlated with the result from the \gls{rsrp} measurement. 
The bigger trees with more dense branches and leaves cause a more significant drop in \gls{rsrp}, and these unreliable links result in decreasing the bitrate and increasing the \gls{rtt}.
Based on the type of foliage and tree, communication speed dropped from 891~Mpbs at the \gls{los} baseline location to 426~Mbps when being behind small trees and further dropped to 323~Mbps when being behind big trees.
Also, the delay is larger for big packets than for small packets, as they may break into multiple Transmission Blocks and be retransmitted multiple times under challenging signal conditions.

\begin{table*}[t]
\centering
\caption{Water effect on \gls{mmwave} propagation, across all the southern sector's \glspl{pci}. Instances with high std are highlighted.}
\begin{tabular}{ccccccccccccc}
\hline
\multicolumn{1}{c}{\multirow{2}{*}{\begin{tabular}[c]{@{}c@{}}Beam\\ type\end{tabular}}} & \multirow{2}{*}{\begin{tabular}[c]{@{}c@{}}Distance\\ (m)\end{tabular}} & \multicolumn{2}{c}{PCI-301} & \multicolumn{2}{c}{PCI-302} & \multicolumn{2}{c}{PCI-303} & \multicolumn{2}{c}{PCI-304} & \multicolumn{3}{c}{TCP} \\
\cline{3-13} 
\multicolumn{1}{c}{} && \multicolumn{1}{c}{\begin{tabular}[c]{@{}c@{}}mean\\ {[}dBm{]} \end{tabular}} & \multicolumn{1}{c}{std} & \multicolumn{1}{c}{\begin{tabular}[c]{@{}c@{}}mean\\ {[}dBm{]} \end{tabular}} & \multicolumn{1}{c}{std} & \multicolumn{1}{c}{\begin{tabular}[c]{@{}c@{}}mean\\ {[}dBm{]} \end{tabular}} & \multicolumn{1}{c}{std} & \multicolumn{1}{c}{\begin{tabular}[c]{@{}c@{}}mean\\ {[}dBm{]} \end{tabular}} & \multicolumn{1}{c}{std} & \multicolumn{1}{c}{\begin{tabular}[c]{@{}c@{}}Bitrate\\ {[}Mbps{]} \end{tabular}} & \multicolumn{1}{c}{\begin{tabular}[c]{@{}c@{}}RTT\\ {[}ms{]} \end{tabular}} & \multicolumn{1}{c}{\begin{tabular}[c]{@{}c@{}}RTT big\\ {[}ms{]} \end{tabular}} \\ \hline
Over water (close to the sea level) & 792 & -103.2 & \textbf{2.33} & -105.2 & \textbf{3.12} & -103.7 & \textbf{2.62} & -105.9 & \textbf{3.35} & 903 & 13.9 & 17.5 \\ \hline
Over water (6m above sea level) & 813 & -111.4 & \textbf{4.61} & -110.6 & \textbf{4.70} & -107.7 & \textbf{3.16} & -105.8 & \textbf{3.25} & 885 & 15.5 & 16.6 \\ \hline
     Ground & 573 & -94.3 & 0.29 & -94.7 & 0.33 & -93.0 & 0.30 & -92.3 & 0.29 \\ \hline
     Ground & 256 & -103.0 & 0.54 & -102.1 & 1.21 & -100.1 & 0.66 & -100.7 & 0.76 \\ \hline
     Ground & 213 & -91.3 & 0.56 & -94.2 & 0.76 & -96.4 & 0.84 & -92.2 & 0.59 \\ \hline
     Ground & 323 & -93.1 & 0.80 & -94.8 & 1.57 & -103.7 & \textbf{3.20} & -97.5 & 1.87 \\ \hline
\end{tabular}
\label{tbl:over_water}
\vspace{-.2cm}
\end{table*}

\begin{figure}[t]
  \centering
  \includegraphics[width=\columnwidth]{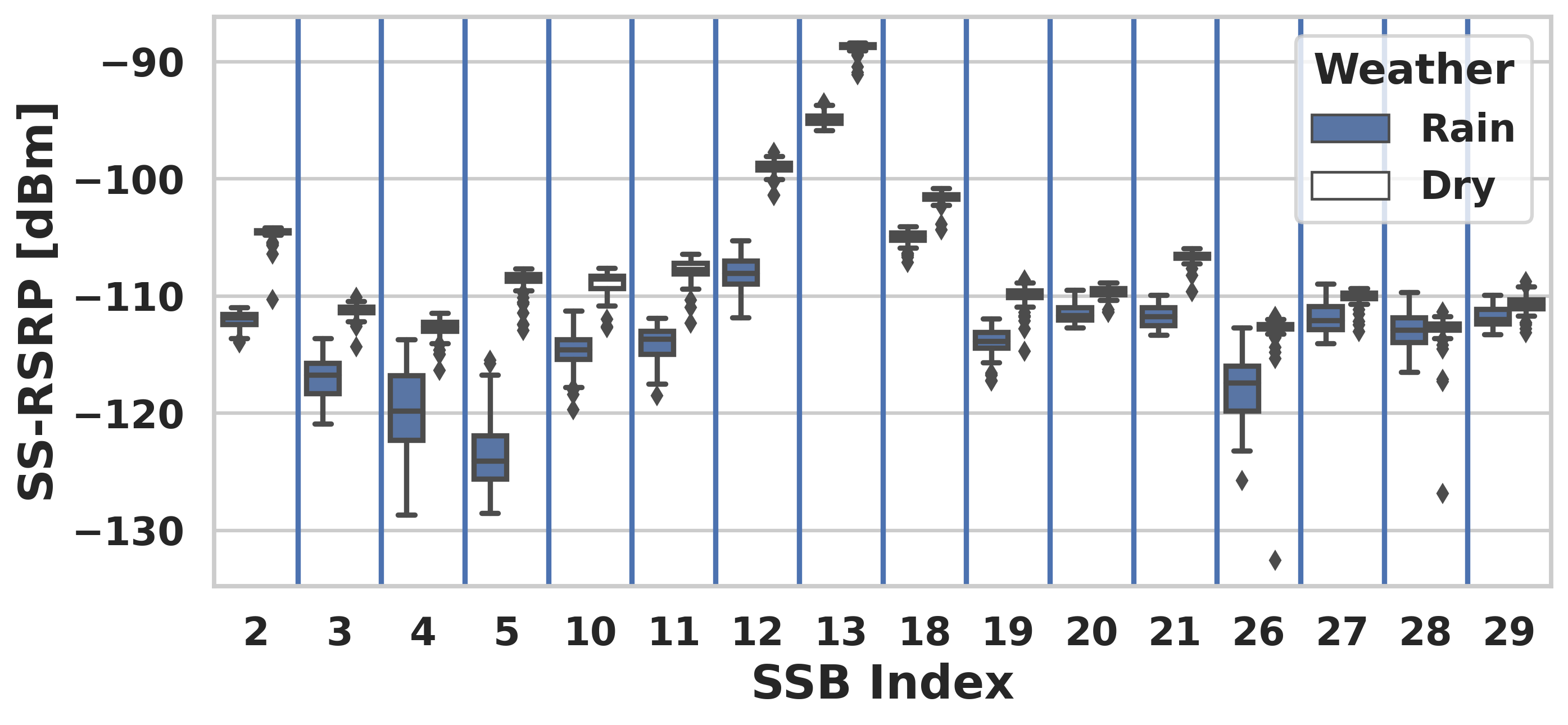}
  \caption{Comparison of received power during rain (left element per boxplot pair) and dry (right element per boxplot pair) weather for a single PCI. Distance from \gls{bs}: 214~meters.}
  \label{fig:rain_volvo_per_ssb}
\end{figure}

\paragraph*{Rain-caused attenuation}
We collected measurements on two different days, with dry and rainy weather, at the same location.
During the rainy day, nearby weather stations reported precipitation between 0.1 and 0.2 mm per minute, which is a moderate to heavy rain intensity.
\Cref{fig:rain_volvo_per_ssb} presents the \gls{rsrp} values measured for different weather conditions for a single \gls{pci}, grouped per beam. Results for the other \glspl{pci} are similar.
As it can be seen from the figure, rain causes a notable drop in the mean \gls{rsrp} and increases its variability in particular for the weaker beams.

The increased variability for the weaker beams is probably caused by multi-path
signal propagation where no single signal component is significantly stronger
than all others.
The most significant components will consist of signals reflected by e.g.
buildings and vegetation, and usually also of the \gls{los} signal.
Strong beams point towards the scanner and the BS antenna gain for the \gls{los}
signal will be high.
Therefore the \gls{los} signal will be much stronger than the reflected signals,
and the received signal power will have little variability.
Weak beams, on the other hand, do not point directly towards the scanner and the
\gls{bs} antenna gain for the \gls{los} signal is therefore low.
In this case the strength of the reflected beams might be comparable to or
larger than the strength of the \gls{los} component.
This results in multi-path fading that gives large variability in the received
signal power.

\paragraph*{Over-water communication}
We also investigate how water surface scattering and reflection affect \gls{mmwave} signals.
This scenario could be relevant when \glspl{mmwave} are used for providing high capacity communications on the shore.
For example, fish farms plan to use \glspl{mmwave} for high definition video communications between fish cages and on-land data centers where advanced signal processing and analytics are used to, e.g., control feeding and monitor fish health~\cite{over_water}.

We performed the measurements at locations with good \gls{los} of the \gls{bs} antenna, on the far side of a small bay.
From the \gls{bs}, the \gls{mmwave} signal first travels over the ground for about 630 meters and then over-water for 160 meters before reaching the scanner.
The measurements were collected at two locations. First, we put the scanner at the shoreline about 50~cm from the water. 
Second, we placed the scanner approximately 6 meters above the water surface.
We could not assess the water wave height directly, but during the measurements, nearby weather stations reported a wind intensity between 4.9~m/s and 5.6~m/s, which typically corresponds to wave heights of 0.6~m to 1.2~m.
The measurement location is relatively sheltered, though, so
we can assume that the wave height is closer to the lower limit of this range.
For comparison, we also made corresponding measurements at \gls{los} locations with different distances to the \gls{bs} where the propagation was solely over-ground (due to terrain and building blockage, we were not able to do over-ground propagation measurements at exactly the same distances as for the over-water measures).
\Cref{tbl:over_water} summarizes our measurements.
The \gls{rsrp} standard deviation for over-water propagation is significantly larger than for over-ground propagation.
The increased standard deviation for the over-water communications can be explained by a model where the received signal consists of a direct \gls{los} component and one or more components (specularly or diffusely) scattered from the water surface.
Since the water surface moves, the scattered components' strength
varies with time, thereby causing signal power variations at the receiver.

Another observation from \Cref{tbl:over_water} is that the difference between the mean \gls{rsrp} values for \glspl{pci} with different frequencies are much larger at 6 meters above the water than at the sea level.
This can be explained by the same propagation model.
When the scanner is located close to the water surface,  the scattered components of the signal are coming from the water immediately in front of the scanner.
In contrast, these components are originated much further out when the scanner is placed far above sea level, so the scattered components' delay compared to the \gls{los} is more variable.
In frequency domain, this translates to flatter fading (i.e., attenuation is almost the same across neighboring frequencies) when the scanner is at sea level than when placed at a higher location.
A consequence of the increased signal variation for over-water reception is that it will be necessary to use a higher link margin in the link budget calculations than when the communication is solely over-ground.
The smaller signal variation for the scanner located at sea level suggests that \gls{mmwave} antennas should be placed close to the water, e.g., on a fish cage.

\Cref{tbl:over_water} also presents the bitrate and \gls{rsrp} values for the over-water transmission. 
The bitrate and the delay at 6 meters above the water are significantly worse than at sea level.
This observation was also expected, as the high-level performance is indeed correlated with the \gls{rsrp} values.

\begin{figure}[t]
  \centering
  \includegraphics[width=0.85\linewidth]{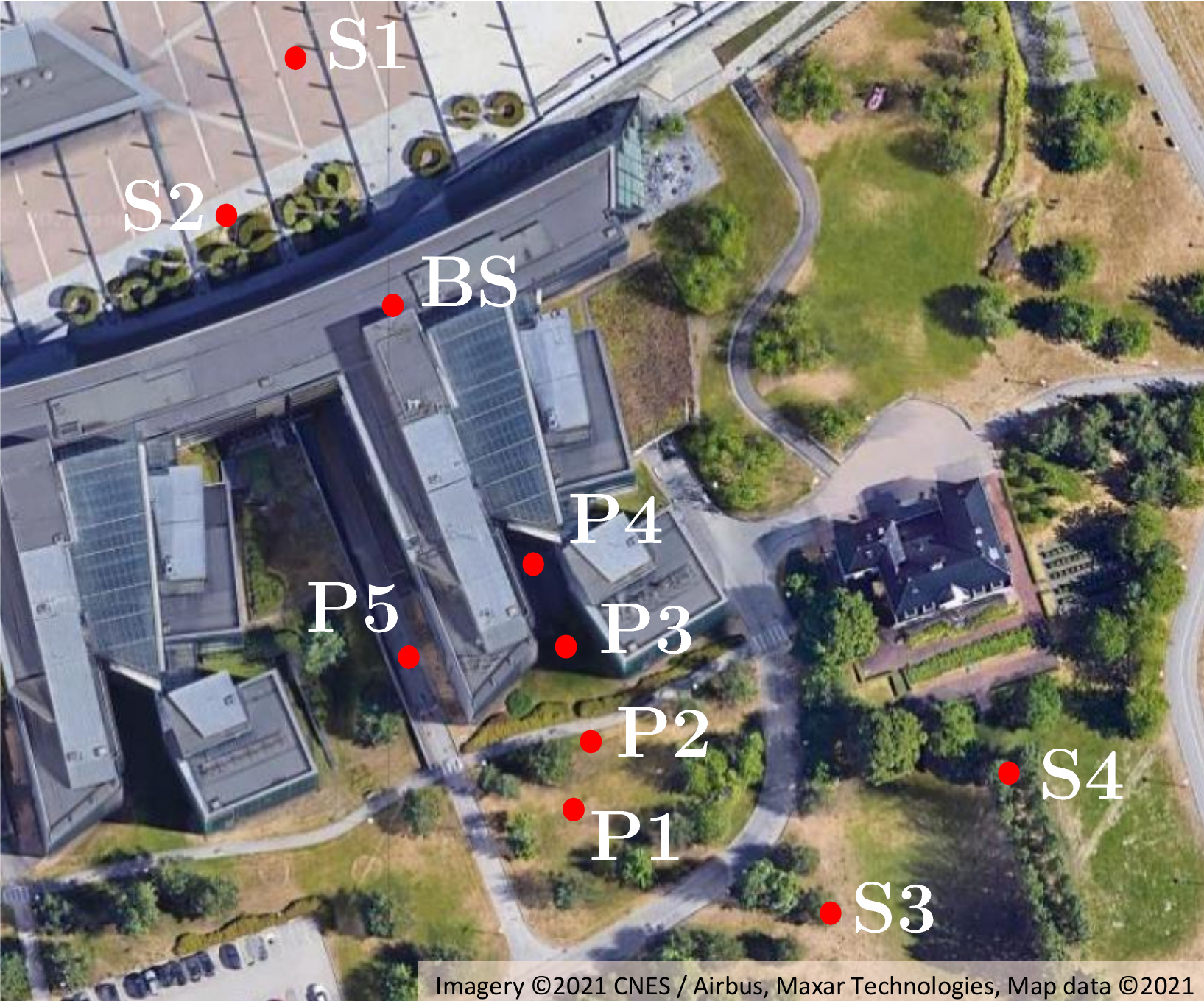}
  \caption{The locations where the nLoS measurements were performed.}
  \label{fig:nlos_map}
\end{figure}

\begin{figure}
\centering
\begin{subfigure}{\linewidth}
  \centering
  \includegraphics[width=\linewidth]{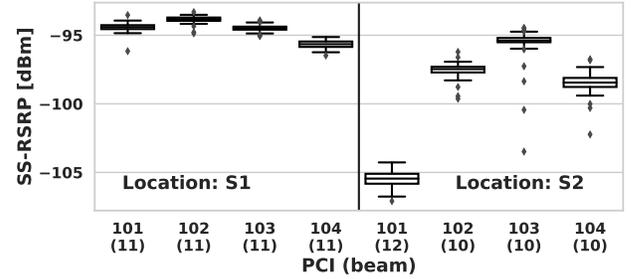}
  \caption{Sector 1 (North)}
  \label{fig:nonLoS_north}
\end{subfigure}
\begin{subfigure}{\linewidth}
  \centering
  \includegraphics[width=\linewidth]{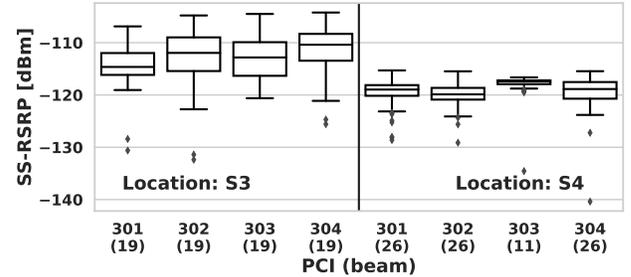}
  \caption{Sector 2 (South)}
  \label{fig:nonLoS_south}
\end{subfigure}
\caption{RSRP at different locations with \gls{nlos} links.
The dominant beam for every PCI is mentioned in parenthesis.}
\label{fig:nlos}
\end{figure}

\paragraph*{nLoS measurement scenarios}
In \gls{mmwave} communications, \gls{los} links can be easily blocked by buildings, moving cars and other obstacles.
In this case, the \gls{rsrp} depends on the \gls{nlos} components, such as reflected, scattered and diffracted waves.
To get an indication of the \gls{nlos} coverage we performed measurements at four locations where the \gls{los} path to the \gls{bs} antenna was blocked by the roof of the building with the \gls{bs} (See \Cref{fig:nlos_map}).
Locations \textbf{S1} and \textbf{S2} in \Cref{fig:nlos_map} are surrounded by high buildings and a large number of rectangular columns placed in a regular pattern in the plaza between the buildings, representing an urban environment. Locations \textbf{S3} and \textbf{S4} are surrounded by trees and some distant buildings, representing a suburban and vegetated environment.
\Cref{fig:nlos} presents the \gls{rsrp} distribution of the best beam (written in parenthesis) for every carrier (\gls{pci}) of the \gls{bs} at these locations.
Even though the orientation of the beams is the same for all the \glspl{pci} at the \gls{bs}, we observe that at locations \textbf{S2} and \textbf{S4} the dominant beam is different for some of the \glspl{pci}.
Further, at location \textbf{S2}, \gls{pci} 101 has degraded performance compared to
the rest of the \glspl{pci}.
Both observations are in contrast to what we measure at \gls{los} locations.
It can be due to the property of different scattering and diffracting materials at different frequencies (i.e., reflection coefficients and penetration losses), but we are unsure about the exact source of these behaviors.

At locations \textbf{S1} and \textbf{S2} the received signal is quite strong, indicating good \gls{nlos} coverage.
This is thought to be due to the rich scattering environment, possible reception
of diffracted (i.e., from the edge of the roof) signal components and the
relatively short distances involved.
At locations \textbf{S3} and \textbf{S4} the received signal level is very low and there is little or no \gls{nlos} coverage.
This is thought to be due to a much poorer scattering environment.
The buildings that might act as good reflectors for the signals are located far
away from both the \gls{bs} antenna and the scanner, hence the signal paths will be very long.
The trees in the surrounding area further attenuate both the reflected and the diffracted signal components.
These results show that in an urban environment, with multiple buildings and scattering elements, the \gls{nlos} components of the \gls{mmwave} signals could compensate for the lack of a \gls{los} link to the \gls{bs}.
In contrast, in vegetated areas, the \gls{los} link is necessary to establish reliable communication.

Finally, we study the end-to-end communication performance over \gls{nlos} 5G \gls{mmwave} links.
To do so, we collect measurements in different locations, close to the buildings, where there is no \gls{los} path to the \gls{bs}. These locations are marked as \textbf{P1}-\textbf{P5} in \Cref{fig:nlos_map}. \Cref{tbl:bitrate_delay_nlos} reports the bitrate and \gls{rtt} values for the mentioned locations.
It is evident that the bitrate sharply drops and the delay increases by moving toward the buildings. 
This is an expected observation as fewer diffracted elements reach the locations close to the building, which results in lower \gls{rsrp} and therefore bitrate.

\begin{table}[t]
\centering
\caption{Summary of the effect of nLoS links to Bitrate and delay.}
\begin{tabular}{llll}
\hline
   Position & Bitrate [Mbps] & RTT [ms] normal size & RTT [ms] big size  \\
\hline
   P1 & 845 & 15.4 & 17.8 \\
   P2 & 573 & 16.5 & 33.9 \\
   P3 & 176 & 15.6 & 22.1 \\
   P4 & 87 &  18.0 & 48.0 \\
   P5 & 177 & 15.4 & 18.1 \\
\hline
\end{tabular}
\label{tbl:bitrate_delay_nlos}
\end{table}

\section{Comparison with Simulations}
\label{sec:simulation}
This section compares our measurement results with simulations based on empirical models to check how accurately statistical models can predict the received signal at different locations.
The complexity of the propagation environment, makes impossible the use of accurate channel models, i.e., ray-tracing based quasi-deterministic channel models~\cite{lecci2020quasi, varshney2021quasi}, to accurately obtain the \gls{nlos}  components of the propagated signal is not possible.
We hence used \gls{3gpp}'s statistical channel model for \gls{mmwave} frequencies~\cite{TSGR2017} to estimate the path loss for different scenarios.
For Sector 1, we use the \gls{umi} and \gls{uma} channel models~\cite{haneda20165g, maccartney2014omnidirectional}, as the \gls{bs} is pointed towards a square surrounded by large buildings and the topology is similar to a typical urban environment, while for Sector 2, we employ the \gls{rma} model~\cite{maccartney2017rural}, as it represents a suburban area with more trees and vegetation and fewer or no buildings.
The omnidirectional path loss model used in our simulations is:

\begin{equation}\label{eq:path_loss}
\text{PL}[dBm] (d) = 20\log_{10}(\frac{4\pi d_0}{\lambda}) + 10n\log_{10}(\frac{d}{d_0})+SF,
\end{equation}
where, $d$ is the distance from transmitter, $\lambda = c/(10^9f)$ [m] is the wavelength, $c=3\times 10^8$ [m/s] is the speed of the light, $SF$ [dB] indicates the shadow fading, whose standard deviation is $\sigma_{SF}$, $n$ represents the path loss exponent and $d_0 = 1$ [m] is the free space reference distance. 
The values for $\sigma_{SF}$ and $n$ at different scenarios are presented in~\Cref{tbl:sim_par}.

\begin{table}[]
\centering
\caption{Simulation parameters.}
\begin{tabular}{lllllll}
\hline
\multirow{2}{*}{Parameter} & \multicolumn{2}{c}{RMa} & \multicolumn{2}{c}{UMa} & \multicolumn{2}{c}{UMi} \\ \cline{2-7} 
& LoS & nLoS & LoS & nLoS & LoS & nLoS \\ \hline
$n$ & 2.3 & 3.1 & 2   & 2.9 & 2   & 3.2 \\
$\sigma_{SF}$    & 1.7 & 6.7 & 4.0 & 7.0 & 4.0 & 7.0 \\ \hline
\end{tabular}
\label{tbl:sim_par}
\end{table}

\begin{figure*}
\centering
\begin{subfigure}{0.45\linewidth}
  \centering
  \includegraphics[width=\linewidth]{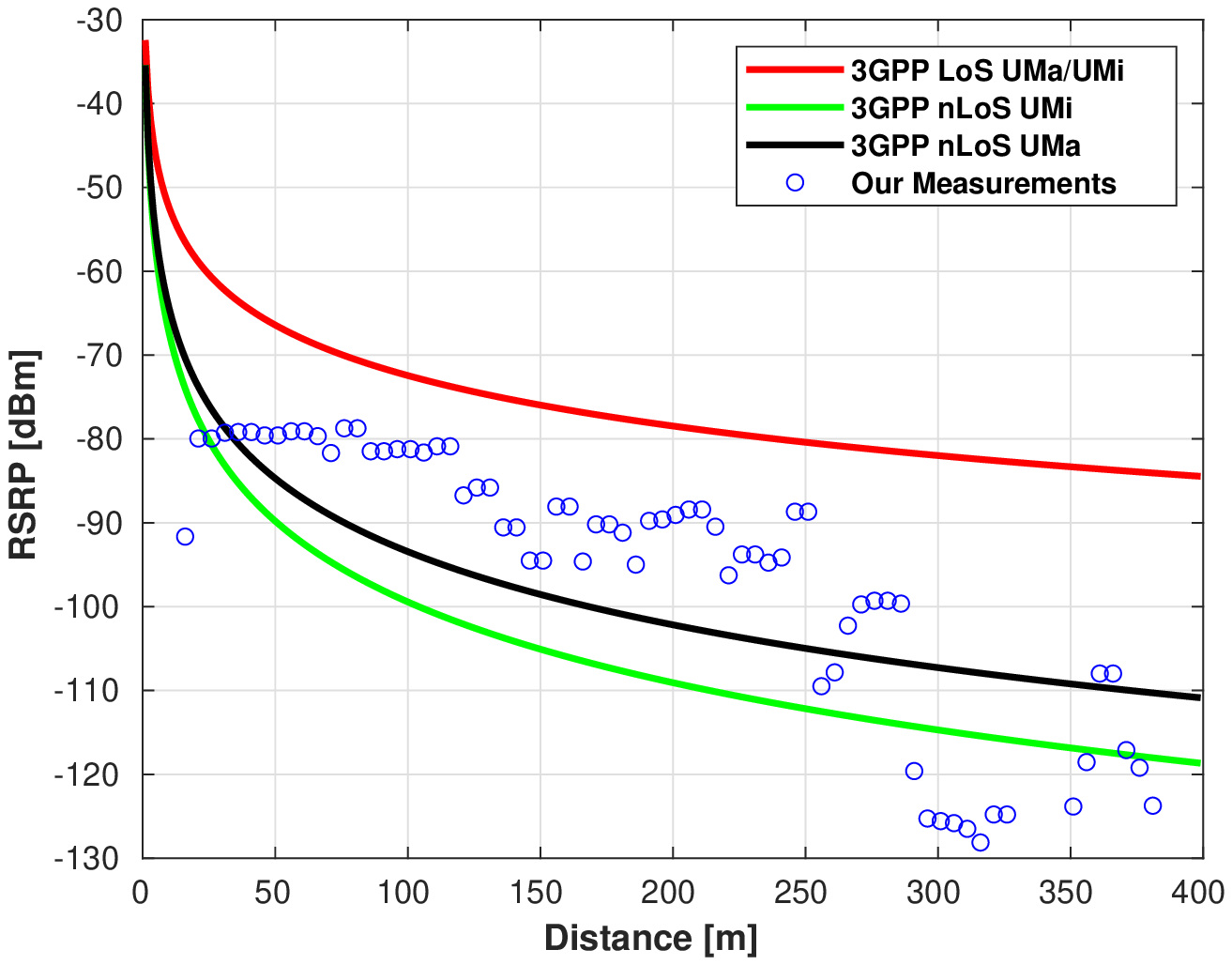}
  \caption{Sector 1}\label{fig:sim1}
\end{subfigure}
\begin{subfigure}{0.45\linewidth}
  \centering
  \includegraphics[width=\linewidth]{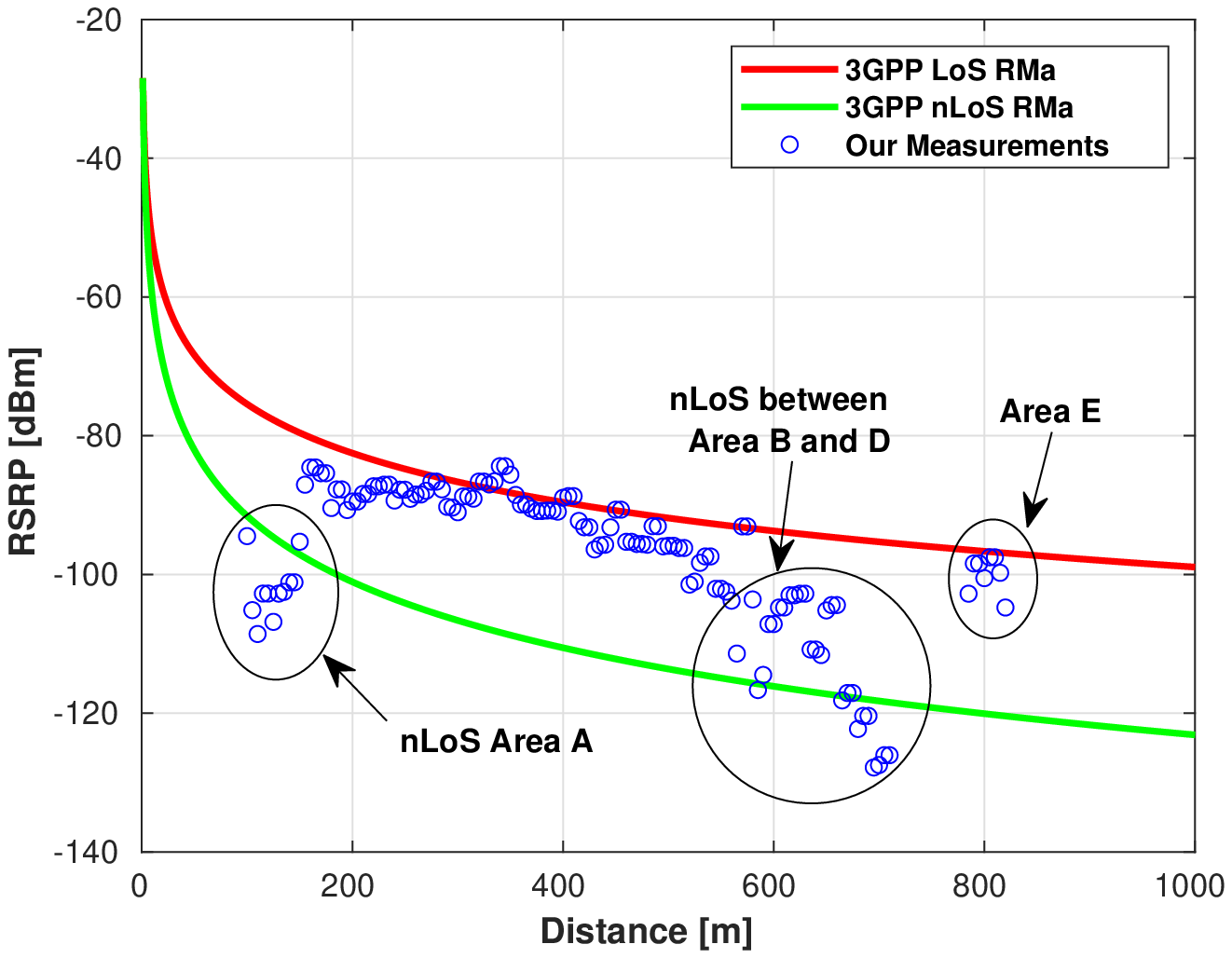}
  \caption{Sector 2}\label{fig:sim2}
\end{subfigure}\par\medskip
\caption{Comparing the maximum measured RSRP with simulated 3GPP models for different distances in (a) Sector 1, and (b) Sector 2.}
\label{fig:simulation}
\vspace{-.2cm}
\end{figure*}

\Cref{fig:simulation} shows the maximum measured \gls{rsrp} of the dominant beam at different distances for Sector 1 and Sector 2 and compares it to the \gls{rsrp} achieved using the omnidirectional path-loss model in \eqref{eq:path_loss}, as is perceived from two omnidirectional isotropic transmit and receive antennas with 0 dBi gain.

The strongest measured \gls{rsrp} in Sector 1 is \textbf{$\approx$}-78~dBm, which is a bit higher than the strongest \gls{rsrp} in the other sector (\textbf{$\approx$}-84~dBm).
As shown in \Cref{fig:sim1}, our measurements can barely fit with the \gls{3gpp} \gls{umi} and \gls{uma} models.
The \gls{rsrp} varies between -78~dBm and and -130~dBm which is always lower than the predicted values by the \gls{los} models and only at larger distance is lower than the \gls{nlos} models.
This figure clearly shows that statistical channel models are not always capable of predicting the \gls{rsrp} accurately in every propagation environment.
\Cref{fig:sim2} shows that the \gls{los} \gls{rma} can estimate the \gls{rsrp} in locations with clear sight to \gls{bs}.
Even in Area \textbf{E}, where the signal propagates overwater, this model predicts the \gls{rsrp} relatively well ($\leq 5$~dBm error).
On the other hand, \gls{nlos} \gls{rma} fails to accurately predict the \gls{rsrp} for \gls{nlos} locations.
The variation in measured power in \gls{nlos} is mainly due to changes in the type of the obstacles, scattering objects, and topography of the environment, rather than pure distance.
Referring to \Cref{fig:coverage_map} and \Cref{fig:nlos_map}, the type and shape of obstacles blocking \gls{los} in area \textbf{A} and the area between \textbf{B} and \textbf{D} are very different and composed of buildings with various shape that highly affect the \gls{rsrp} in these areas.

\begin{table}[t]
\centering
\caption{Comparing the measured RSRP [dBm] with simulations for different scenarios.}
\begin{tabular}{lllll}
\hline
\multirow{2}{*}{Scenario} & \multirow{2}{*}{\begin{tabular}[c]{@{}c@{}}Distance\\ {[}m{]}\end{tabular}} & \multirow{2}{*}{Measurement} & \multicolumn{2}{c}{Simulation} \\ \cline{4-5} &&& Omni & Strongest \\

\hline
  RMa LoS & 160 & -84.5 & -80.8 & -31.5 \\
  RMa LoS & 250 & -87.7 & -85.2 & -36.0 \\
  RMa LoS & 350 & -85.6 & -88.8 & -39.6 \\
  RMa LoS & 450 & -90.6 & -91.4 & -42.2 \\
  RMa LoS & 570 & -93.0 & -94.0 & -44.8 \\
  RMa LoS & 810 & -97.5 & -97.7 & -48.4 \\ \hline
  RMa nLoS &  80 & -103.9 &  -90.4 & -41.2 \\
  RMa nLoS & 120 & -102.7 &  -93.4 & -44.2 \\
  RMa nLoS & 590 & -114.4 & -117.6 & -68.4 \\
  RMa nLoS & 690 & -120.3 & -117.5 & -68.3 \\ \hline
  UMa LoS &  60 & -79.0 & -64.5 & -17.2 \\
  UMa LoS & 160 & -88.0 & -73.4 & -26.1 \\
  UMa LoS & 250 & -88.6 & -77.0 & -29.6 \\
  UMa LoS & 500 & -97.8 & -83.2 & -35.9 \\ \hline
  UMa nLoS &   15 &  -91.6 &  -63.7 & -16.6 \\
  UMa nLoS &  300 & -114.3 & -100.5 & -53.4 \\
  UMa nLoS &  450 & -123.6 & -106.0 & -58.8 \\
 \hline
\end{tabular}
\label{tbl:measure_vs_simulation}
\end{table}

\Cref{tbl:measure_vs_simulation} compares the \gls{rsrp} recorded during the measurement campaign against the \gls{3gpp} channel models which are used to estimate the \gls{rsrp}, both on case of omnidirectional and directional antenna gain patterns at transmitter and receiver.
To obtain the values for the strongest \gls{rsrp}, we use the NYUSIM simulator~\cite{sun2017novel}, which searches for the best pointing angle among all possible pointing angles employing the specified antenna details (i.e., azimuth and elevations of receiver and transmitter antennas) in both transmitter and receiver. 
The details of the procedure is out of the scope of the this paper and interested readers may refer to \cite{sun2017novel, ju2019millimeter} and the references therein for more details.
For this simulation, we assumed that \gls{bs} and \gls{ue} are equipped with 16 and 4 uniform linear array antenna elements, respectively.

As seen in~\Cref{tbl:measure_vs_simulation} and \Cref{fig:simulation}, only in limited \gls{los} scenarios, the omnidirectional path loss models perform well in predicting the \gls{mmwave} channels.
In complicated environments, especially in \gls{nlos} locations with many complex scattering objects, using more advanced ray-tracing-based models is difficult, if not impossible, so some extent of measurement is needed to estimate the \gls{rsrp} accurately.
It is also seen that in the best case, the received power at different locations is as good as the omnidirectional power.
The gap between the measurement (and omnidirectional power) with the Strongest power is significant for all scenarios ($\approx 50$~dBm). 
This vast gap reveals how much an adaptive dynamic beamformer can improve performance.
This specifically makes more sense in some \gls{mmwave} use-cases where the \glspl{ue} are stationary, i.e., \gls{fwa} or do not move fast.
Some codebook-based beamforming techniques can be employed for these use-cases where the optimal transmitter/receivers antenna configuration can be learned and saved for future uses.

\section{Conclusion}
\label{sec:result_summary}
Based on our measurement campaign results, we can confirm that the range of \gls{mmwave} cells is strongly affected by the presence of obstacles, such as buildings, trees, or human bodies, as already observed in the literature. 
As expected, the \gls{los} beam is generally the most robust in all situations.

For what concerns the effect of foliage, we observed a degradation of the signal quality when the \gls{los} is obstructed by trees, particularly when moved by (even light) wind. 
However, this generally affects all beams, so that the dominant beam remains the same. 
The measurements have also revealed that wide water surfaces, especially in presence of waves, can generate time-varying scattering phenomena that affect the stability of the received signal.
This is pronounced if the receiver is higher than the water surface and thus collects more water-reflected waves. 
The propagation of \gls{mmwave} on water surfaces is, hence, critical and would require further investigation to determine the limitations of links involving floating stations.
Comparing the measured \gls{rsrp} with the simulated omnidirectional (based on \gls{3gpp}'s path loss models) and strongest \gls{rsrp}  (optimal antenna configuration) shows the necessity of measurement in especially complex \gls{nlos} environment, in predicting cell coverage and recognizing coverage holes. It also reveals the considerable improvement that \glspl{mmwave} can achieve by employing a perfect beamformer.

In conclusion, we can claim that our empirical analysis of a commercial 5G cellular system has confirmed the validity of previous studies carried out on prototypal or \gls{poc} deployments or via simulations. Clearly, general theoretical models cannot perfectly capture the complexity of real installments, and some significant deviations from model predictions and real-world measurements have been observed in certain cases. This will require the implementation of self-tuning capabilities on commercial installations to adapt the BS configuration to the specificities of the environments, though a first, rough performance estimate can be done using the theoretical models. The effects of body blockage, rain, and trees on the propagation of mmWave signals previously reported in the literature have been mostly observed also in our study. This confirms that the commercial version of the mmWave communication interface does not show any significant limitation with respect to pre-commercial versions, which was not guaranteed. On the other hand, it does not bring any improvement either.
Finally, we noticed that beam selection in a real commercial setting is likely less critical than feared, since the strongest beam appears to remain rather stable in time and space, also in bad weather conditions, so that beam swiping techniques should be able to track the best beam direction rather easily.

The above observations should help interested stakeholders make more informed decisions when deploying 5G solutions utilizing the \gls{mmwave} spectrum.
Our next steps involve testing verticals' use cases at the same cell site to check if their service requirements are met.

\ifCLASSOPTIONcaptionsoff
  \newpage
\fi

\bibliographystyle{IEEEtran}
\bibliography{mmWave}

\end{document}